\documentclass{article}

\usepackage[preprint,nonatbib]{aaj2026}
\usepackage[authoryear,round,sort]{natbib}
\bibpunct[, ]{(}{)}{;}{a}{}{,}

\usepackage[utf8]{inputenc}
\usepackage[T1]{fontenc}
\usepackage{hyperref}
\usepackage{url}
\usepackage{booktabs}
\usepackage{amsfonts}
\usepackage{amsmath}
\usepackage{mathtools}    
\usepackage{amssymb}
\usepackage{nicefrac}
\usepackage{microtype}
\usepackage{xcolor}
\usepackage[normalem]{ulem}   
\usepackage{graphicx}

\graphicspath{{./}{figures/}}

\makeatletter
\renewcommand{\@noticestring}{%
  Published in \textit{Georisk: Assessment and Management of Risk for
  Engineered Systems and Geohazards}, Taylor \& Francis.
  \url{https://doi.org/10.1080/17499518.2026.2707500}%
}
\makeatother

\title{TabPFN Extensions for Interpretable Geotechnical Modelling}

\author{%
  Taiga Saito$^{a}$, Yu Otake$^{a}$, Daijiro Mizutani$^{a}$, Stephen Wu$^{b,c}$ \\[0.5em]
  $^{a}$Department of Civil and Environmental Engineering, Tohoku University, \\
  6-6-06 Aramaki Aoba, Aoba-ku, Sendai, Miyagi 980-8579, Japan \\[0.2em]
  $^{b}$Research Organization of Information and Systems, The Institute of Statistical Mathematics \\
  10-3 Midori-cho, Tachikawa, Tokyo 190-8562, Japan \\[0.2em]
  $^{c}$Department of Statistical Science, The Graduate University for Advanced Studies \\
  10-3 Midori-cho, Tachikawa, Tokyo 190-8562, Japan \\[0.4em]
  \texttt{taiga.saito.r3@dc.tohoku.ac.jp}
}

\begin{document}

\maketitle

\begin{abstract}
Geotechnical site characterisation relies on sparse, heterogeneous
borehole data, where uncertainty quantification and interpretability
matter as much as predictive accuracy. We evaluate
TabPFN~\citep{Hollmann2025}, a tabular foundation model, and its
\texttt{tabpfn-extensions} library~\citep{TabPFNExtensionsGitHub} on two
geotechnical tasks:
(1) soil-type classification from N-value and shear-wave velocity data
as a controlled illustrative case, and (2) iterative imputation of five
mechanical parameters ($s_\mathrm{u}$, $E_{\mathrm{u}}$,
${\sigma'}_\mathrm{p}$, $C_\mathrm{c}$, $C_\mathrm{v}$) in
BM/AirportSoilProperties/2/2025~\citep{Otake2025}. Without retraining,
we apply cosine-similarity analysis to TabPFN embeddings, visualise
predictive distributions, and compute SHAP attributions. On
the regression benchmark we compare TabPFN with mean imputation, linear
regression, random forests, XGBoost, and HBM; introduce a proxy
decomposition of predictive uncertainty across context-perturbation
classes; and propagate marginal $C_\mathrm{c}$ and ${\sigma'}_\mathrm{p}$
distributions through a one-dimensional consolidation model to obtain
the reliability index $\beta$ and serviceability exceedance probability
$P_\mathrm{f}$. Embeddings exhibit label-consistent Clay/Sand grouping;
iterative imputation reduces RMSE for all five targets, with TabPFN
lowest on four; SHAP attributions are consistent with the Skempton
compression-index correlation and the inverse
preconsolidation-pressure-water-content dependence; the within-posterior
component is largest in the proxy decomposition. We position the
contribution as a worked evaluation workflow that may complement
established methods for data-scarce geotechnics, not as algorithmic
innovation.
\end{abstract}


\section{Introduction}

Geotechnical site characterisation is inherently challenged by the scarcity
and heterogeneity of borehole measurement data.
Reliable prediction of soil mechanical properties---undrained shear strength
($s_\mathrm{u}$, from unconfined compression test), undrained secant modulus
($E_{\mathrm{u}}$), preconsolidation stress (${\sigma'}_\mathrm{p}$), compression index
($C_\mathrm{c}$), and coefficient of consolidation ($C_\mathrm{v}$)---is
essential for foundation design, liquefaction assessment, and ground improvement
planning, yet field data often consist of only tens to hundreds of samples.
In engineering applications, the objective extends beyond predictive accuracy:
model predictions---particularly for safety-critical decisions such as
foundation design---must be supported by interpretable justification consistent
with established physical understanding, and prediction uncertainty must be
quantified to enable reliability-based assessments.
Consequently, purely black-box approaches that offer high accuracy without
interpretability remain difficult to adopt in geotechnical practice.

In geotechnical engineering, parameter inference has traditionally relied on
empirical correlations and regression models derived from laboratory and
field observations.
More recently, probabilistic and Bayesian approaches have been proposed to
quantify uncertainty in geotechnical parameters and to leverage data from
multiple sites.
For example, hierarchical Bayesian models have been developed for
constructing quasi-site-specific multivariate probability
distributions~\citep{Ching2021, Wu2022}, with related developments
including quasi-site-specific probabilistic modelling of sparse,
incomplete, and three-dimensional spatially varying soil
data~\citep{Ching2022QSS}, and sparse-measurement reconstruction of
three-dimensional subsurface properties~\citep{Huang2025Tucker}, and
similarity-based site retrieval
methods have been proposed using Bayesian
measures~\citep{Sharma2022, Sharma2023}.
Data-driven approaches---including dictionary learning for transformation
models~\citep{Cai2025}, clustering-based regionalisation~\citep{Cai2024},
and high-order dependence modelling~\citep{Saito2025MIDM}---have further
expanded the methodological toolkit for site characterisation.
\citet{Phoon2022} provided a comprehensive review of the challenges inherent
in data-driven site characterisation, and \citet{Phoon2025FR} framed the
problem of quantifying site uniqueness as a ``facial recognition'' challenge
for geotechnical data.
Despite these advances, the integration of predictive accuracy,
interpretability, and uncertainty quantification within a single,
easy-to-deploy framework remains an open challenge.

Recent reviews of machine learning in geotechnics~\citep{PhoonZhang2023FoML}
have highlighted data-centricity, the prevalence of small and incomplete
datasets, and the need for interpretability and uncertainty quantification
as outstanding challenges in bringing machine-learning methods into
geotechnical practice.
The emergence of foundation models has opened new possibilities for inference
in engineering problems.
These models can be viewed as a generalisation of conventional statistical
inference approaches: rather than training a task-specific model from scratch,
a foundation model acquires broad predictive capabilities from large-scale
pretraining and adapts to new tasks through in-context learning, providing
flexible predictions without extensive model training or hyperparameter
tuning.
TabPFN~\citep{Hollmann2025} is a transformer-based tabular foundation model
trained via meta-learning on prior data synthesised from a diverse set of
causal relationships commonly observed in tabular data.
It performs in-context learning: given a small observed dataset at inference
time, the model
conditions its predictions on the entire context without gradient-based
fine-tuning.
Throughout this paper, \emph{training data} refers to the observed samples
provided as context to TabPFN at inference time---not data used to update
model parameters---and \emph{test data} denotes the samples for which
predictions are sought.
In a companion study~\citep{Saito2025}, we demonstrated that TabPFN
achieves competitive predictive accuracy for geotechnical site
characterisation on the BM/AirportSoilProperties benchmark, rivalling
conventional regression approaches even under severely limited training data.
The \texttt{tabpfn-extensions} library~\citep{TabPFNExtensionsGitHub} further exposes
the model's internal representations, providing access to learned embeddings
and SHAP-based feature attribution without any model retraining.
The base TabPFN model itself natively outputs full posterior distributions over
predictions, which we exploit for iterative inference and uncertainty quantification.

From this perspective, the present paper adopts the generic extension tools
of TabPFN~\citep{Hollmann2025,TabPFNExtensionsGitHub} for the specific task
of site characterisation in order to perform exploratory analyses and to
investigate how foundation-model-based approaches may support interpretable
parameter inference in geotechnical engineering.
Rather than proposing a new algorithm or methodological innovation in
the strict sense, this study examines the potential of
applying an existing foundation model framework to geotechnical problems,
with emphasis on interpretability, uncertainty quantification, and
reliability-based decision support.
The contribution is positioned as an exploratory proof-of-concept and
a worked evaluation workflow, situated alongside---and complementary to---the
quasi-site-specific inference framework~\citep{Phoon2022,Phoon2025FR}
and the hierarchical Bayesian and dictionary-learning
frameworks~\citep{Ching2021,Wu2022,Cai2024,Cai2025}.
Specifically, we address five aspects:
\begin{enumerate}
  \item We apply embedding-based diagnostics in two settings: cosine-similarity
        analysis for a controlled soil classification case, and a
        target-specific embedding diagnostic for borehole-level grouping in
        the regression benchmark. These analyses examine whether TabPFN's
        internal representations group samples consistently with constructed
        soil-type labels or site-conditioned structure, while remaining
        diagnostic rather than directly validated imputation mechanisms.
  \item We perform iterative imputation of missing mechanical parameters
        and visualise full posterior distributions, providing
        sample-level uncertainty estimates 
        that we then couple explicitly with reliability-based reasoning.
  \item We compute SHAP-based feature importance, revealing a 
        two-regime structure
        in which mechanical parameters dominate the SHAP attribution for
        all five mechanical targets, while index properties contribute
        secondarily.
  \item We benchmark TabPFN against mean imputation, linear regression,
        random forests, XGBoost, and the hierarchical Bayesian model (HBM)
        under the same iterative-imputation framework on
        the BM/AirportSoilProperties benchmark~\citep{Otake2025}.
  \item We introduce a proxy decomposition of TabPFN's predictive
        uncertainty and an illustrative settlement-reliability
        calculation that propagates marginal predictive distributions of $C_\mathrm{c}$
        and ${\sigma'}_\mathrm{p}$ through a one-dimensional consolidation
        model to obtain $\beta$ and $P_\mathrm{f}$.
\end{enumerate}

\section{Classification Problem: Embedding-Based Similarity Analysis}
\label{sec:classification}

Site similarity quantification is central to geotechnical site
characterisation~\citep{Sharma2022, Sharma2023}.
This section uses the controlled Clay/Sand case to inspect whether
TabPFN embeddings---extracted as a byproduct of inference---can provide
a visual diagnostic of sample-level similarity. The purpose is to
illustrate the workflow, not to establish an operational coverage
assessment method for realistic geotechnical classification data.

\subsection{Dataset and Setup}

\begin{figure}[htbp]
  \centering
  \includegraphics[width=\linewidth]{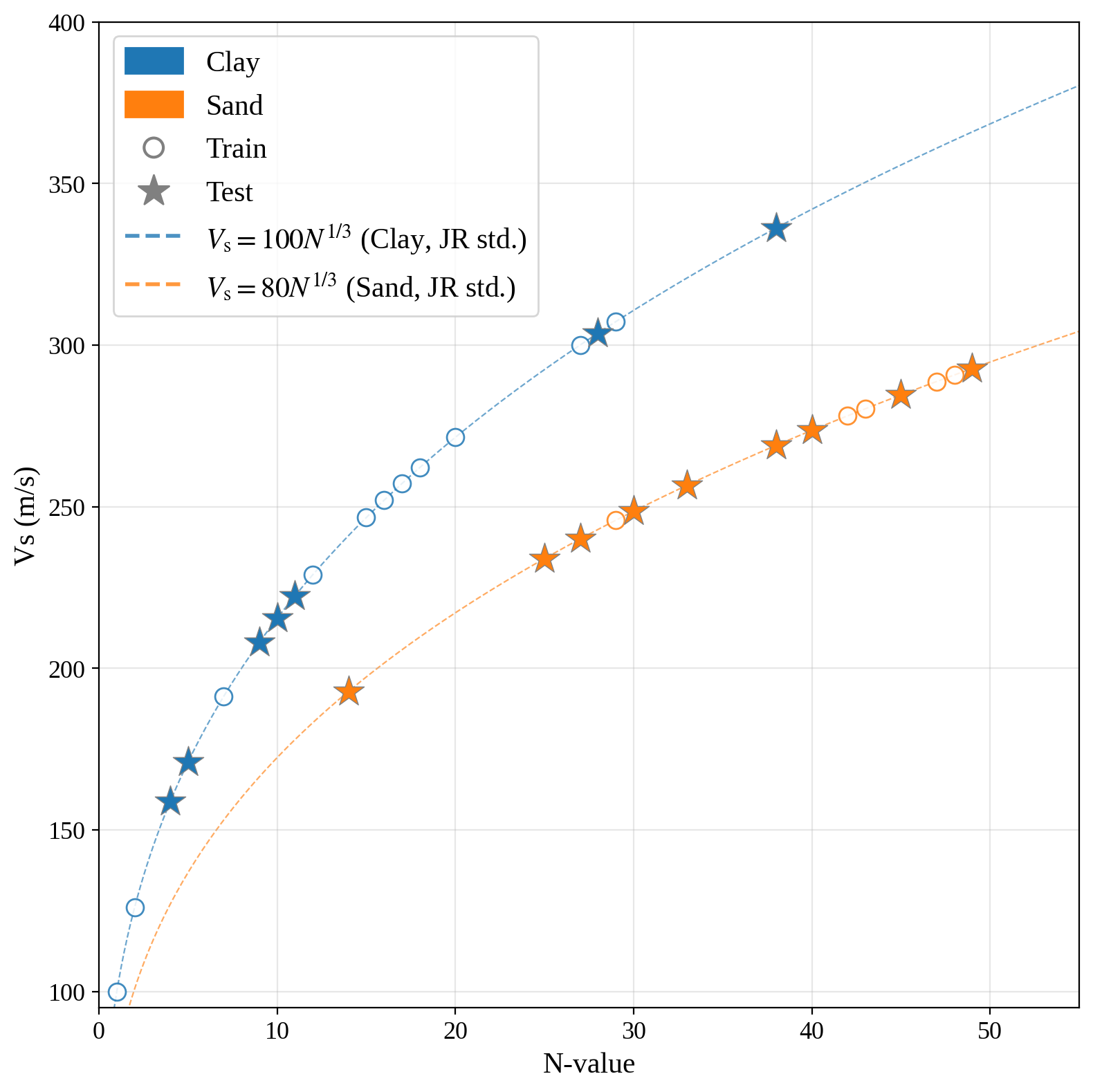}
  \caption{Scatter plots of training (circles) and test (stars) data in the
           $N$\,--\,$V_\mathrm{s}$ space, coloured by soil class
           (Clay: blue, Sand: orange).
           Training and test sets each contain 16 samples.
           Reference curves $V_\mathrm{s} = 100N^{1/3}$ and
           $V_\mathrm{s} = 80N^{1/3}$ from the Japanese railway
           seismic design standard~\citep{RTRI2012} are overlaid;
           Clay samples follow the upper curve and Sand samples
           the lower curve.}
  \label{fig:clus_scatters}
\end{figure}

We use a synthetic geotechnical dataset consisting of borehole samples
labelled as Clay or Sand.
Input features are the standard penetration test N-value and shear-wave
velocity $V_\mathrm{s}$.
The dataset is synthetic in the sense that $V_\mathrm{s}$ values were not
independently measured but derived from borehole N-values via the empirical
formulae $V_\mathrm{s} = 100N^{1/3}$ (Clay) and $V_\mathrm{s} = 80N^{1/3}$ (Sand)
prescribed in the Japanese railway seismic design standard~\citep{RTRI2012},
making $V_\mathrm{s}$ a deterministic function of $N$ for each soil class.
The Clay/Sand labels themselves come from the original field
record and are independent of $N$. However, the per-class
$V_\mathrm{s}$ assignment above is label-conditioned, so the perfect
separability noted below should be read as a property of this data
construction rather than as evidence that the model has uncovered an
intrinsic class structure.
Because $V_\mathrm{s}$ is uniquely determined by $(N, \text{soil class})$,
the two classes are perfectly separable in the $(N, V_\mathrm{s})$ space,
making the classification task itself straightforward; the primary
purpose of this section is therefore to illustrate the interpretability
of TabPFN's learned embeddings and predicted probability surfaces,
rather than to benchmark classification accuracy.
The dataset was split into training and test sets; the training set was used
to condition TabPFN's in-context predictions, and the test set was held out
for evaluation.
The dataset comprises 32 samples in total, randomly split equally into
16 training samples (11 Clay, 5 Sand) and 16 test samples
(7 Clay, 9 Sand) (Table~\ref{tab:rtri_data}).
This sample size is consistent with the small-data regime typical of
geotechnical site investigations, where borehole costs limit the
available data to a few tens of samples.
Notably, the training set contains only five Sand samples, all with
$N \geq 29$, whereas the test set includes Sand samples at $N$ as low
as 14---a region not covered by any Sand training sample.
Combined with the 11:5 Clay-to-Sand imbalance, this configuration
provides a challenging inference scenario in which the model must
extrapolate beyond the observed Sand training distribution.
As shown in Figure~\ref{fig:clus_scatters}, the two soil classes are
well-separated in the $N$--$V_\mathrm{s}$ space.

\begin{table}[htbp]
  \caption{All samples in the synthetic geotechnical dataset.
           No.\ corresponds to the axis labels in
           Figure~\ref{fig:embedding_heatmap}: training samples are
           ordered along the row axis and test samples along the
           column axis (Clay first, sorted by ascending $N$-value;
           Sand second, sorted by ascending $N$-value).}
  \label{tab:rtri_data}
  \centering
  \footnotesize
  \begin{tabular}{rrrr@{\hspace{2em}}rrrr}
    \toprule
    \multicolumn{4}{c}{Training set ($n = 16$)} &
    \multicolumn{4}{c}{Test set ($n = 16$)} \\
    \cmidrule(r){1-4}\cmidrule(l){5-8}
    No. & $N$ & $V_\mathrm{s}$ (m/s) & Soil &
    No. & $N$ & $V_\mathrm{s}$ (m/s) & Soil \\
    \midrule
     1 &  1 & 100.000 & Clay &  1 &  4 & 158.740 & Clay \\
     2 &  2 & 125.992 & Clay &  2 &  5 & 170.998 & Clay \\
     3 &  7 & 191.293 & Clay &  3 &  9 & 208.008 & Clay \\
     4 & 12 & 228.943 & Clay &  4 & 10 & 215.443 & Clay \\
     5 & 15 & 246.621 & Clay &  5 & 11 & 222.398 & Clay \\
     6 & 16 & 251.984 & Clay &  6 & 28 & 303.659 & Clay \\
     7 & 17 & 257.128 & Clay &  7 & 38 & 336.198 & Clay \\
     8 & 18 & 262.074 & Clay &  8 & 14 & 192.811 & Sand \\
     9 & 20 & 271.442 & Clay &  9 & 25 & 233.921 & Sand \\
    10 & 27 & 300.000 & Clay & 10 & 27 & 240.000 & Sand \\
    11 & 29 & 307.232 & Clay & 11 & 30 & 248.579 & Sand \\
    12 & 29 & 245.785 & Sand & 12 & 33 & 256.603 & Sand \\
    13 & 42 & 278.082 & Sand & 13 & 38 & 268.958 & Sand \\
    14 & 43 & 280.272 & Sand & 14 & 40 & 273.596 & Sand \\
    15 & 47 & 288.706 & Sand & 15 & 45 & 284.551 & Sand \\
    16 & 48 & 290.739 & Sand & 16 & 49 & 292.744 & Sand \\
    \bottomrule
  \end{tabular}
\end{table}

\subsection{Classification Results}
\label{sec:classification_results}

TabPFN is applied as an in-context classifier: the 16 training samples are
provided as context, and the model predicts the class probability for each
test sample without any gradient-based fine-tuning or hyperparameter tuning.
On this test split, all 16 test samples were correctly classified
(accuracy = 1.00, ROC-AUC = 1.00).
Rather than focusing on this accuracy result alone, the following analysis
examines the model's predicted probability surface to compare the inferred
boundary with the reference curves and to identify data-sparse regions.

\begin{figure}[htbp]
  \centering
  \includegraphics[width=\linewidth]{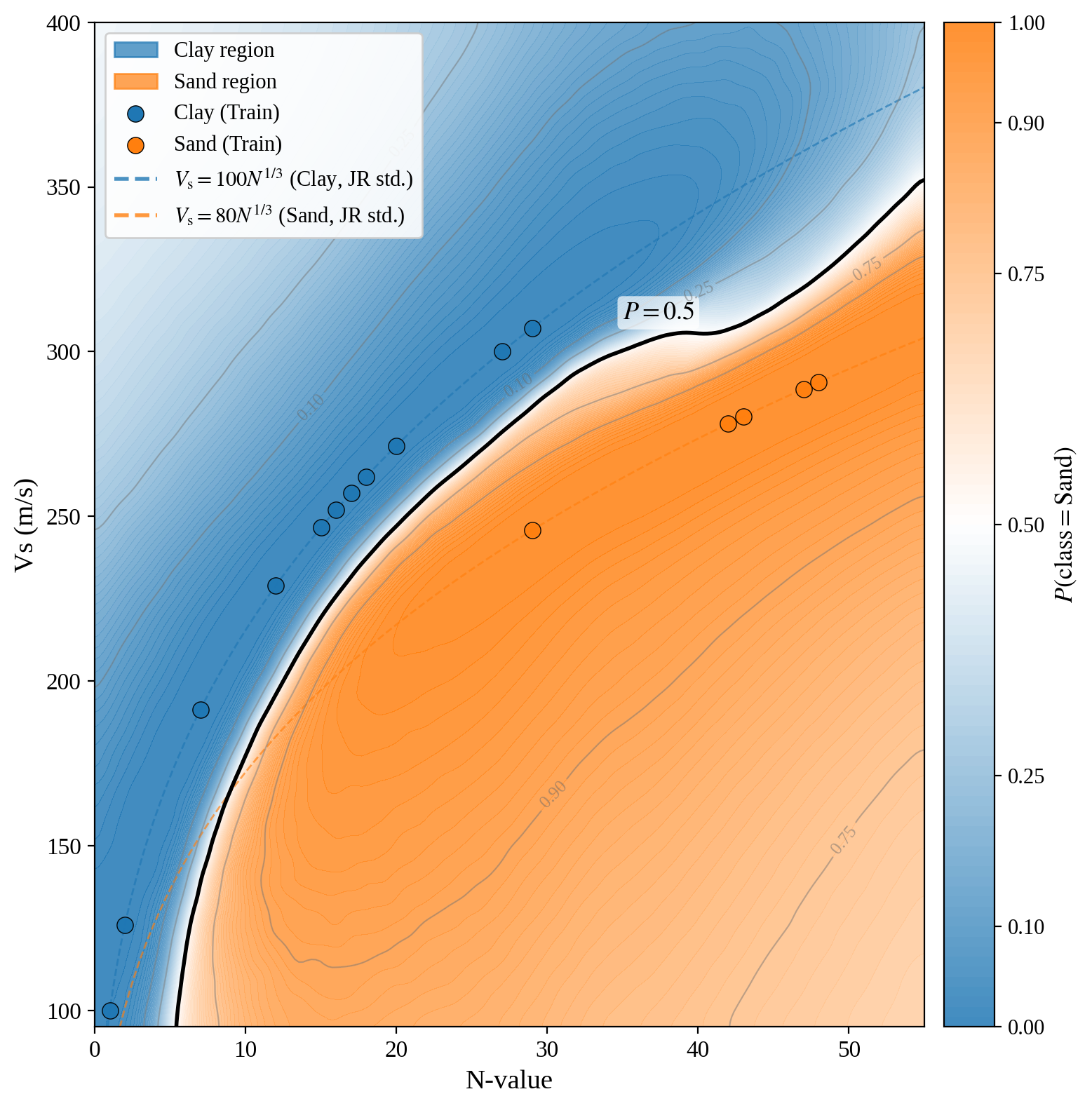}
  \caption{Predicted probability of Sand class $P(\text{Sand})$ over the
           $N$\,--\,$V_\mathrm{s}$ domain, with the training samples
           overlaid (circles).
           The bold contour marks the decision boundary at $P(\text{Sand}) = 0.5$;
           grey contours indicate the 0.1, 0.25, 0.75, and 0.9 levels.}
  \label{fig:proba_heatmap}
\end{figure}

Figure~\ref{fig:proba_heatmap} shows the predicted probability of Sand,
$P(\text{Sand})$, over the $N$--$V_\mathrm{s}$ feature space.
The model assigns high-confidence predictions near the training samples:
the deep blue region coincides with the cluster of Clay training points,
and the deep orange region aligns with the Sand training points at
$N \geq 29$.
In this constructed feature space, the decision boundary broadly follows
the separation induced by the two reference curves rather than revealing
an independently validated soil-class mechanism. In feature-space corners
far from both class contexts---high $N$ combined with very low
$V_\mathrm{s}$ (lower right of the plot), and very low $N$ combined with
high $V_\mathrm{s}$ (upper left)---the predicted probability moves toward
$P \approx 0.5$, indicating lower confidence under extrapolation. We
therefore read the probability surface as a diagnostic of how TabPFN
interpolates within this synthetic setup.
The perfect test accuracy on the held-out split should be read with
this caveat in mind. The test set does not probe the lower-left region
near or below the Sand reference curve; in this region, the TabPFN
$P=0.5$ boundary dips below $V_\mathrm{s}=80N^{1/3}$ for $N \leq 9$
and remains close to it for $N=10$--$12$. A hypothetical Sand sample on
the reference curve in this corner would therefore lie near the decision
boundary, where the prediction confidence is low rather than reliably
correct. Test accuracy alone is therefore an incomplete summary of the
model's behaviour in this data-sparse region.

The predictive probabilities provide a complementary signal even in
the absence of misclassified test samples. The most uncertain test
sample, at $(N=14, V_\mathrm{s}=193)$, has
$P(\mathrm{Sand})=0.898$ and predictive entropy $0.475$, compared with
$0.071$ for the next-highest test sample. Within Euclidean radius $50$
in $(N,V_\mathrm{s})$, this sample has zero Sand and two Clay training
neighbours, consistent with its elevated entropy.

\subsection{Embedding Analysis}
\label{sec:embedding}

A distinctive property of TabPFN as a foundation model is its contextually
enriched internal representations, made accessible through the
\texttt{tabpfn-extensions} library.
TabPFN encodes each sample into a high-dimensional latent vector
(embedding) via multi-head attention over the entire training context,
so that each vector captures not only the sample's own feature values
but also how it relates to every other sample in the dataset.
This contextual encoding---in which each embedding is shaped by the
full training set rather than derived from a single sample in
isolation---
makes the embedding dependent on the training context and therefore
potentially useful for inspecting sample relationships. We use pairwise
cosine similarity as one summary of inter-sample affinity in the learned
representation space, while recognising that this summary is diagnostic
rather than a physical distance metric.

\begin{figure}[htbp]
  \centering
  \includegraphics[width=0.85\linewidth]{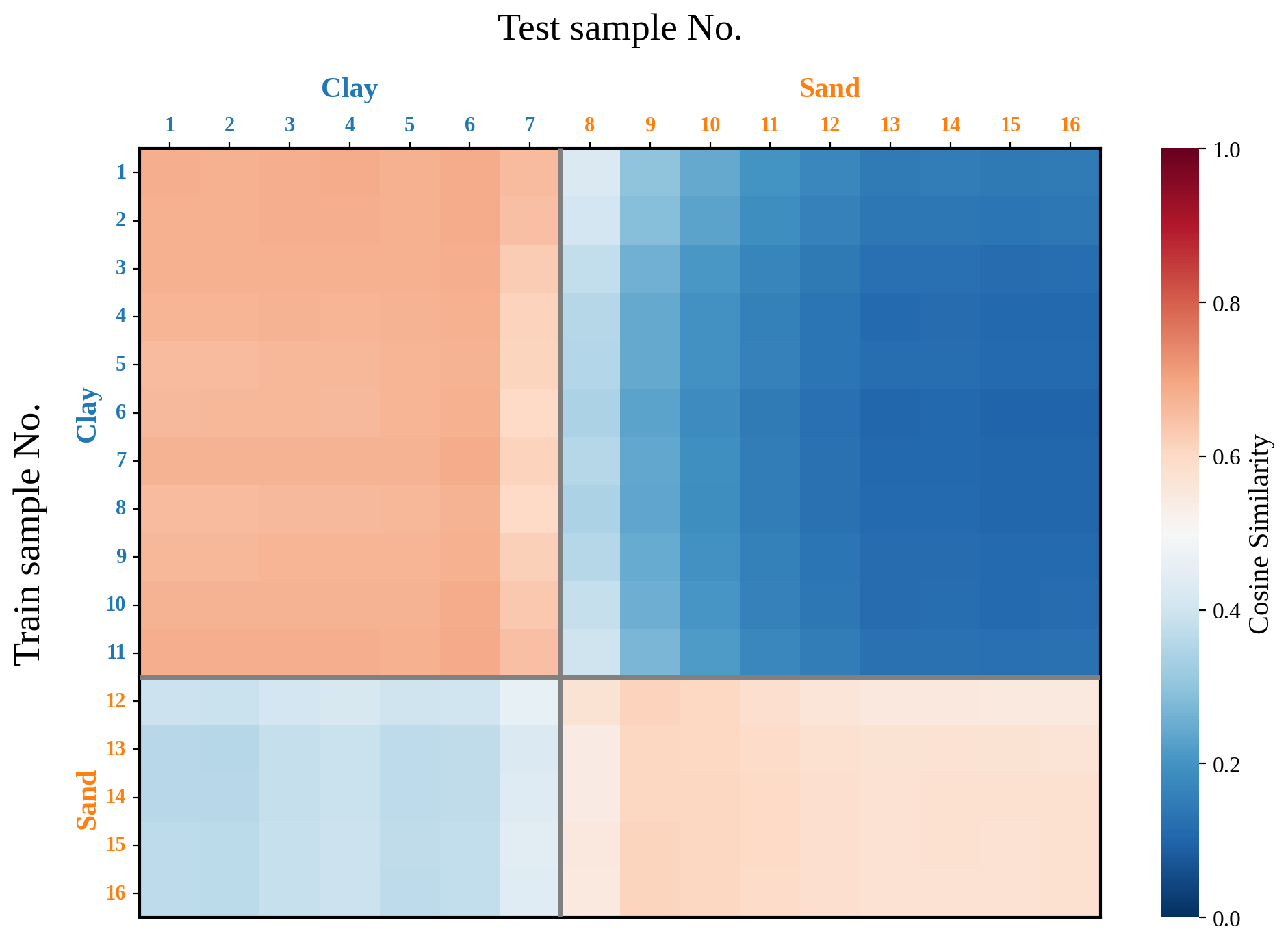}
  \caption{Cosine similarity heatmap of TabPFN embeddings between
           test and training samples (axis labels correspond to Table~\ref{tab:rtri_data}
           No.).
           The block-diagonal structure indicates that, in this controlled
           case, test embeddings are more similar to training embeddings
           from the same constructed class.
           Test sample No.\,8 (first Sand test sample; $N = 14$,
           $V_\mathrm{s} = 193$\,m/s) shows visibly lower similarity
           with the training Sand block,
           illustrating how the heatmap can flag a boundary sample.}
  \label{fig:embedding_heatmap}
\end{figure}

Figure~\ref{fig:embedding_heatmap} shows the cosine-similarity matrix
between test embeddings (columns, No.\ 1--16) and training embeddings
(rows, No.\ 1--16), ordered by soil class (Clay first, then Sand) and
ascending $N$-value within each class, consistent with the sample
numbering in Table~\ref{tab:rtri_data}.

The block-diagonal structure in Figure~\ref{fig:embedding_heatmap}
indicates that Clay and Sand samples are grouped in the embedding space
for this constructed dataset. We interpret this as label-consistent
grouping in a controlled illustrative case.

A noteworthy exception is Test sample No.\,8 (the first Sand test
sample; $N = 14$, $V_\mathrm{s} = 193$\,m/s), whose column in the
heatmap exhibits a conspicuously lower cosine similarity with the
training Sand block compared to other Sand test samples, producing
near-neutral values across both soil-class blocks.
Test No.\,8 falls at $N = 14$, entirely outside the range of Sand
training data ($N \geq 29$), and its $(N, V_\mathrm{s})$ coordinates
overlap with the Clay training distribution.
The model thus lacks a clear analogy in its training context for this
sample, and the embedding-level ambiguity is a direct reflection of
that geometric uncertainty.
This ambiguity is only partly visible from the predicted probability:
Test No.\,8 is still classified correctly, but its predictive entropy is
elevated relative to the other test samples. The cosine-similarity
heatmap provides a complementary diagnostic by showing that the sample
has weaker affinity to the Sand training block than the other Sand test
samples.

The cosine-similarity matrix 
provides a quantitative summary of how
similar each test sample is to the training samples in this controlled
illustrative case. Whether the diagnostic transfers to realistic data
with transformation uncertainty and class overlap is left to future work.

This example demonstrates how TabPFN embeddings can provide additional
interpretability beyond classification accuracy by revealing the internal
structure of the dataset---offering a data-driven analogue to the
similarity-based site characterisation methods established in the
geotechnical literature~\citep{Sharma2022, Sharma2023}.
These results suggest that embedding representations derived from foundation
models may serve as a quantitative basis for inter-site similarity assessment,
complementing conventional index-based approaches in geotechnical practice.

\section{Regression Problem: Interpretable Multivariate Inference}
\label{sec:regression}

This section 
examines how TabPFN and its extensions can be used for
interpretable multivariate inference for geotechnical parameters.
The iterative inference procedure and posterior distributions
(Sections~\ref{sec:iteration}--\ref{sec:posterior}) rely on TabPFN's native
prediction capabilities, whereas 
SHAP-based feature importance (Section~\ref{sec:shap}) and
the embedding-based borehole-structure diagnostic
(Section~\ref{sec:embedding_diagnostic}) use the permutation explainer and
the contextual-embedding extraction provided by
\texttt{tabpfn-extensions}.

\subsection{Problem Setting}

We address benchmark problem BM/AirportSoilProperties/2/2025~\citep{Otake2025},
which involves the prediction of five mechanical parameters---undrained shear
strength $s_\mathrm{u}$, undrained secant modulus $E_{\mathrm{u}}$, preconsolidation
pressure ${\sigma'}_\mathrm{p}$, compression index $C_\mathrm{c}$, and coefficient
of consolidation $C_\mathrm{v}$---from six index properties: saturation
degree $S_r$, unit weight $\gamma_t$, void ratio $e$, liquid limit $LL$,
plastic limit $PL$, and natural water content $w$.
This problem is one of several benchmarks developed to compare
machine-learning approaches to data-driven site characterisation,
following the framework of the benchmark suite introduced
by~\citet{Phoon2022Benchmarks}.
The dataset comprises $n_\mathrm{train} = 2{,}766$ training samples and
$n_\mathrm{test} = 20$ test samples, each representing test results obtained
from a soil specimen sampled at a given depth of a borehole.
The training samples contain no missing values; for the test samples, some
mechanical parameters may already be observed while others are treated as
missing targets to be predicted. Following the benchmark
terminology of \citet{Otake2025}, we refer to the $2{,}766$ complete
benchmark training rows as the Big Indirect Data (BID) context for this
task---a generic, multi-site database that provides the context for
site-specific inference at the $20$ test samples.
Unlike \citet{Saito2025}, who used the borehole identifier and the
six index properties as the feature set ($d = 7$, with separate models
fitted for each (missing-pattern, target) cell), the present study
additionally treats the four other mechanical parameters as
iteratively-imputed covariates ($d = 11$ per target prediction),
enabling inter-parameter information to be exploited during inference.

\subsection{Iterative Inference Procedure}
\label{sec:iteration}

To leverage the cross-parameter dependencies among the five mechanical
parameters, we adopt an iterated conditional mean procedure---a
deterministic analogue of Gibbs sampling in which each mechanical parameter
is sequentially updated using the posterior mean rather than a random
draw from the conditional distribution.
The procedure is as follows:
\begin{enumerate}
  \item \textbf{Initialisation.}
        For each test sample, mechanical parameters that are already
        observed are kept at their measured values; the remaining
        (missing) parameters are initialised with the training-set
        mean.
        The borehole identifier and the six index properties
        ($S_r$, $\gamma_t$, $e$, $LL$, $PL$, $w$) are treated as fixed
        observed inputs ($d_\mathrm{fixed} = 7$: one grouping identifier
        plus six index properties).
  \item \textbf{Iterative update} (repeat for $k = 1, \ldots, 10$).
        For each target parameter $j \in \{s_\mathrm{u}, E_{\mathrm{u}},
        {\sigma'}_\mathrm{p}, C_\mathrm{c}, C_\mathrm{v}\}$:
        \begin{enumerate}
          \item 
                Construct the feature matrix from the borehole identifier,
                the six index properties, and the current estimates of the four
                non-target mechanical parameters (10 physical covariates plus
                one borehole identifier; $d = 11$ total inputs).
          \item Fit TabPFN on the training data and predict the full
                posterior distribution of parameter~$j$ for all test
                samples.
          \item For test samples in which parameter~$j$ is missing,
                update its estimate with the posterior mean;
                observed values are held fixed.
        \end{enumerate}
  \item \textbf{Output.}
        Retain the posterior distributions from the final iteration for
        uncertainty quantification and downstream analysis.
\end{enumerate}
This pattern-agnostic strategy estimates each parameter column-wise for
all test samples uniformly, without conditioning on the specific
combination of observed and missing parameters for each sample---in
contrast to \citet{Saito2025}, who fitted separate models for each
distinct missing-data pattern.

\begin{figure}[htbp]
  \centering
  \includegraphics[width=0.85\linewidth]{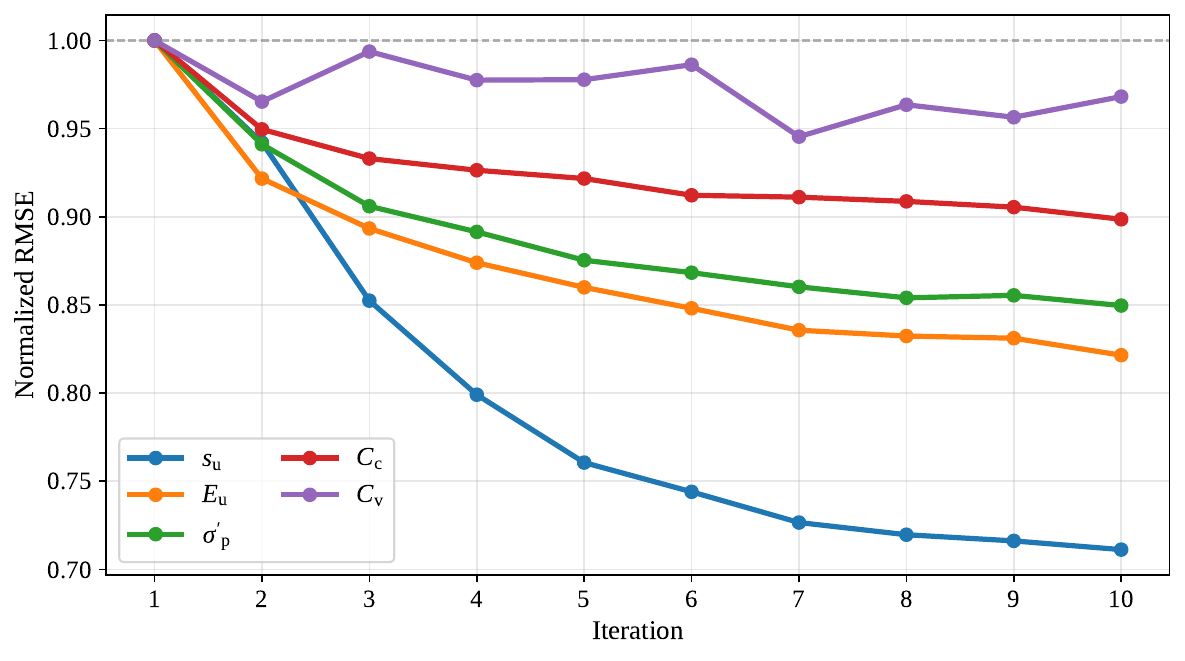}
  \caption{Normalised RMSE (RMSE / RMSE at iteration~1) per iteration for
           each mechanical parameter. Values below 1.0 indicate improvement
           over the initial estimate.}
  \label{fig:rmse_trend}
\end{figure}

Figure~\ref{fig:rmse_trend} shows the normalised RMSE (relative to
iteration~1) for each mechanical parameter across iterations.
Final RMSE decreased for all five targets relative to iteration~1,
although the trajectories were not strictly monotonic for
$C_\mathrm{v}$ and ${\sigma'}_\mathrm{p}$. The largest relative
improvement is observed for $s_\mathrm{u}$
(${\sim}29\%$ reduction, $13.08 \to 9.30$~kN/m$^2$), followed by
$E_{\mathrm{u}}$ (${\sim}18\%$), ${\sigma'}_\mathrm{p}$
(${\sim}15\%$), and $C_\mathrm{c}$ (${\sim}10\%$);
$C_\mathrm{v}$ exhibits only a marginal ${\sim}3\%$ reduction
($35.72 \to 34.58$~m$^2$/s, with iteration-wise oscillations between
$33.77$ and $35.72$), consistent with the iterative scheme
propagating inter-parameter information for most parameters but
contributing little for $C_\mathrm{v}$.
This is attributable to the high intrinsic variability of $C_\mathrm{v}$,
which limits the useful information that correlated parameters can
contribute.
Although the number of iterations ($K = 10$) was set without a formal
convergence criterion, the final-iteration RMSE reduction observed for all
five parameters suggests that iterative conditioning can be useful for this
benchmark; establishing optimal stopping rules remains a topic for future
investigation.

Table~\ref{tab:rmse} summarises the absolute RMSE values at iteration~1
and iteration~10.
The improvement is most pronounced for $s_\mathrm{u}$ and
$E_{\mathrm{u}}$, while $C_\mathrm{c}$ and $C_\mathrm{v}$ show more
modest reductions.

\begin{table}[htbp]
  \caption{RMSE at iteration~1 and iteration~10 for each mechanical
           parameter.}
  \label{tab:rmse}
  \centering
  \begin{tabular}{lrr}
    \toprule
    Parameter & RMSE (iter.\ 1) & RMSE (iter.\ 10) \\
    \midrule
    $s_\mathrm{u}$\,(kN/m$^2$)    & 13.08 & 9.30 \\
    $E_{\mathrm{u}}$\,(kN/m$^2$)           & 3277  & 2692  \\
    ${\sigma'}_\mathrm{p}$\,(kN/m$^2$)    & 30.47 & 25.89 \\
    $C_\mathrm{c}$\,(---)          & 0.118 & 0.106 \\
    $C_\mathrm{v}$\,(m$^2$/s)    & 35.72 & 34.58 \\
    \bottomrule
  \end{tabular}
\end{table}

\subsection{Posterior Distributions}
\label{sec:posterior}

\begin{figure}[htbp]
  \centering
  \includegraphics[width=\linewidth]{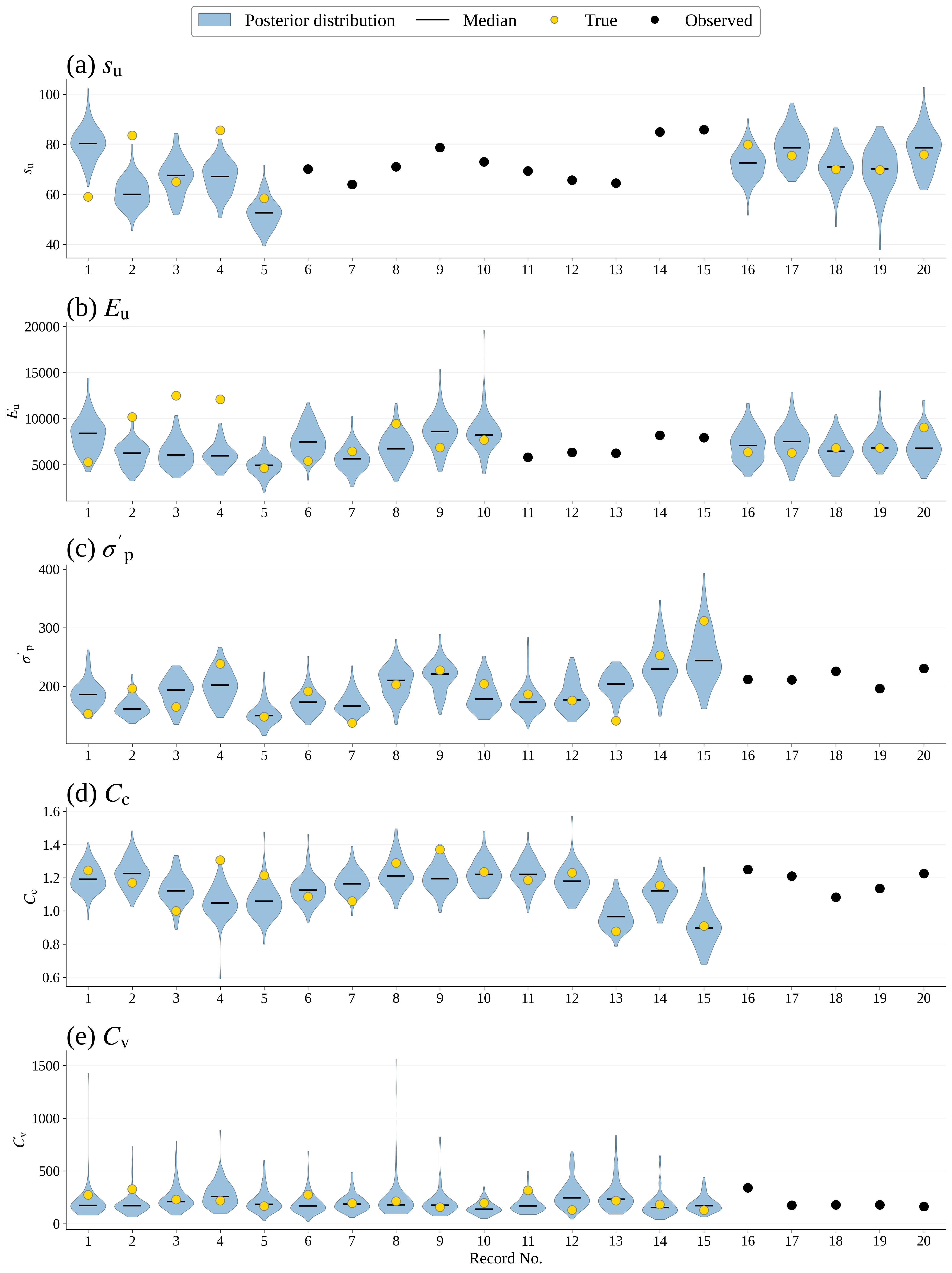}
  \caption{Posterior distributions at iteration~10 for all five mechanical
           parameters across test samples.
           Violin width represents the probability density;
           black bar = median; yellow circle = true value;
           black filled circle = observed value (parameter was not missing
           for that sample and therefore not predicted).}
  \label{fig:violin}
\end{figure}

Figure~\ref{fig:violin} displays violin plots of the posterior distributions
at the final iteration (iteration~10) for all five mechanical parameters
across test samples.
Each violin represents the full predictive distribution for a single
sample; the black bar indicates the median, and the yellow circle the true value.
Black filled circles indicate samples for which the parameter was already
observed (not missing) and therefore not predicted.

The posterior widths reflect parameter-specific predictability: $C_\mathrm{c}$
and ${\sigma'}_\mathrm{p}$ show relatively narrow distributions, while
$C_\mathrm{v}$ exhibits broad posteriors consistent with its high intrinsic
variability.
In most samples, the true value lies within the high-density region of the
posterior, suggesting that the model's uncertainty estimates are physically
plausible.
One practical caveat is that TabPFN's discretised output representation can
occasionally assign non-negligible probability mass to extreme values,
resulting in elongated distribution tails; the median is therefore the
more reliable point estimate in such cases.

\subsection{Feature Importance via SHAP}
\label{sec:shap}

\begin{figure}[htbp]
  \centering
  \includegraphics[width=\linewidth]{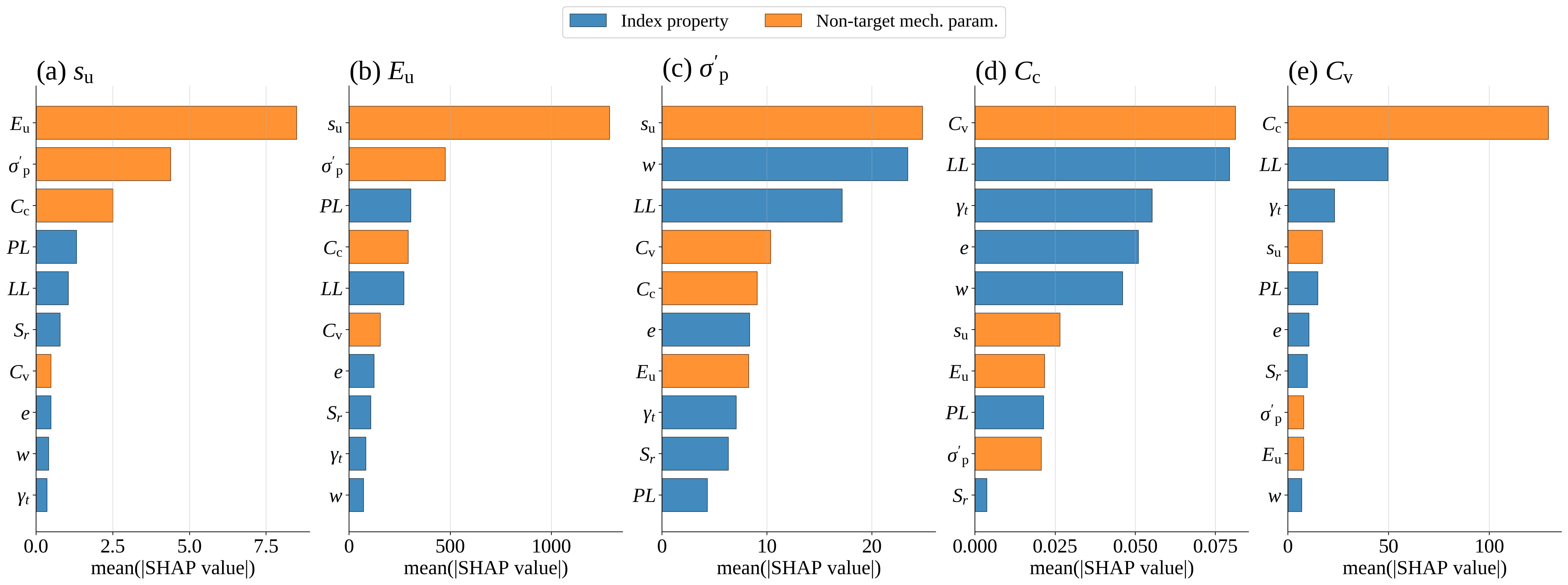}
  \caption{Mean absolute SHAP values for each input feature across all test
           samples, shown separately for each mechanical parameter.
           Blue bars: index properties; orange bars: non-target mechanical
           parameters used as features in the iterative imputation scheme.}
  \label{fig:shap_bar}
\end{figure}

\begin{figure}[htbp]
  \centering
  \includegraphics[width=\linewidth]{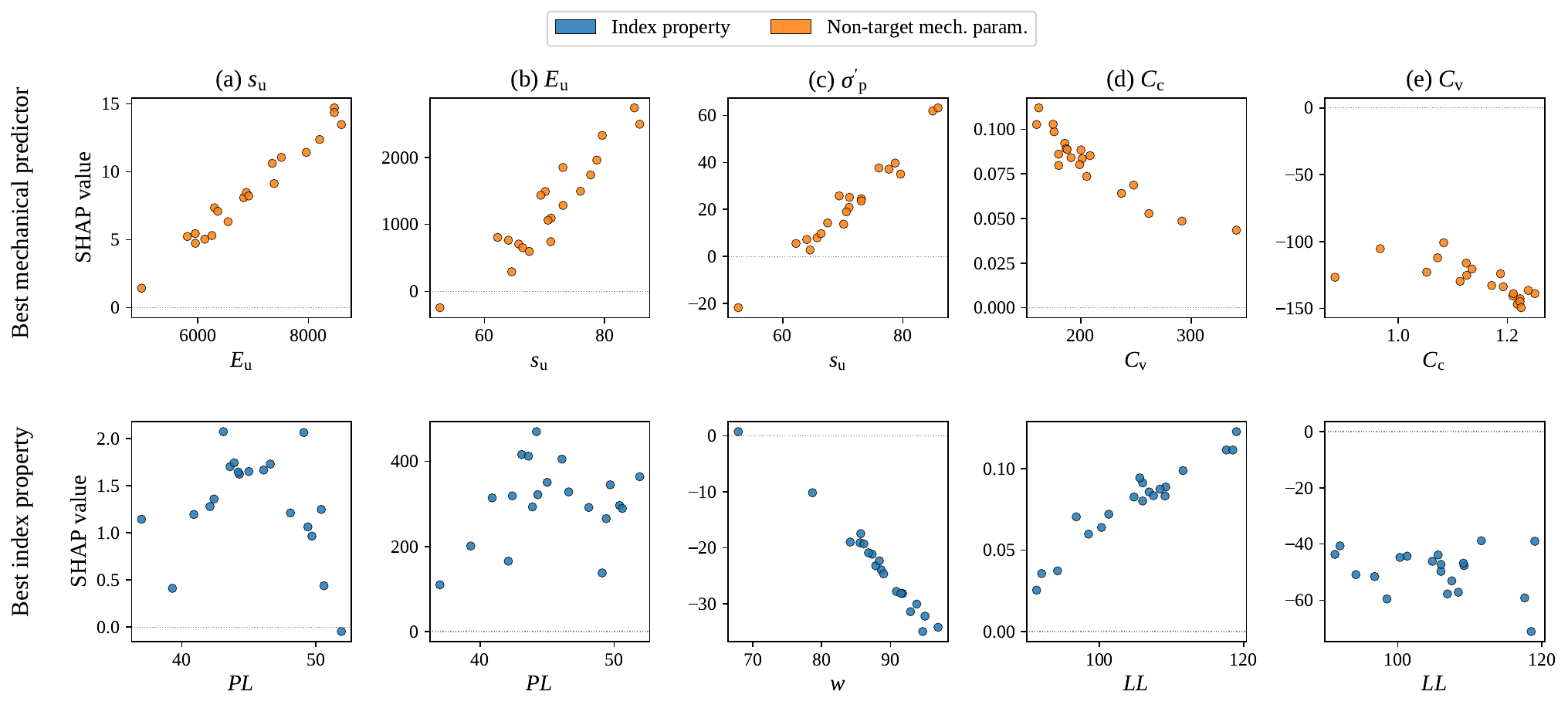}
  \caption{
           SHAP value versus feature value for the dominant predictors
           of each mechanical target. Columns (a)--(e) correspond to
           $s_\mathrm{u}$, $E_{\mathrm{u}}$, ${\sigma'}_\mathrm{p}$,
           $C_\mathrm{c}$, and $C_\mathrm{v}$, respectively. The upper
           row shows the best non-target mechanical predictor per target
           by mean $|\mathrm{SHAP}|$
           ($s_\mathrm{u}\!\leftarrow\!E_{\mathrm{u}}$,
           $E_{\mathrm{u}}\!\leftarrow\!s_\mathrm{u}$,
           ${\sigma'}_\mathrm{p}\!\leftarrow\!s_\mathrm{u}$,
           $C_\mathrm{c}\!\leftarrow\!C_\mathrm{v}$,
           $C_\mathrm{v}\!\leftarrow\!C_\mathrm{c}$); the lower row
           shows the best index-property predictor per target
           ($s_\mathrm{u}, E_{\mathrm{u}}\!\leftarrow\!PL$;
           ${\sigma'}_\mathrm{p}\!\leftarrow\!w$;
           $C_\mathrm{c}, C_\mathrm{v}\!\leftarrow\!LL$).
           Positive SHAP values indicate that the feature increases the
           predicted value; negative values indicate a decrease.}
  \label{fig:shap_scatter}
\end{figure}

A key objective of this section is to inspect the inter-parameter
attribution pattern associated with TabPFN's iterative inference procedure.
Prior Bayesian approaches to geotechnical multivariate inference---such as
hierarchical Bayesian models~\citep{Ching2021, Wu2022}---characterise
inter-parameter correlations explicitly through a probabilistic model.
SHAP-based attribution provides a complementary, model-agnostic perspective:
by quantifying each feature's marginal contribution to individual predictions,
it allows the dependency structure to be read directly from the trained model
without assuming a parametric correlation form.
To identify which input features most strongly influence each mechanical
parameter prediction, we computed SHAP values~\citep{Lundberg2017} using the permutation explainer
provided by \texttt{tabpfn-extensions}, 
applied to target-specific TabPFN diagnostic models that use 10 physical covariates
(the six index properties plus the four non-target mechanical
parameters at their final-iteration imputed values), with the
borehole identifier withheld so that the attribution focuses on
index properties and non-target mechanical parameters rather than
on site-identifier effects.
Figure~\ref{fig:shap_bar} shows the mean absolute SHAP value for each feature
grouped by parameter, and Figure~\ref{fig:shap_scatter} plots the SHAP value
of the 
best mechanical predictor
and the best index-property predictor against its feature value for each
target parameter.
The upper row shows the strongest non-target mechanical predictor
for each target, while the lower row shows the strongest index-property
predictor; for all five mechanical targets, the all-feature dominant
predictor is another mechanical parameter rather than an index property.

The results reveal a consistent first-order attribution pattern:
for all five mechanical targets, the largest mean $|\mathrm{SHAP}|$
is assigned to another mechanical parameter rather than to an index
property. This should be interpreted as an associative feature-attribution
result under the iterative imputation scheme, rather than as evidence
of causal dominance. Concretely:
\begin{itemize}
  \item $s_\mathrm{u}$ is paired with $E_{\mathrm{u}}$ in a reciprocal
        attribution: $s_\mathrm{u}\!\leftarrow\!E_{\mathrm{u}}$ and
        $E_{\mathrm{u}}\!\leftarrow\!s_\mathrm{u}$, consistent with their
        well-known empirical coupling in fine-grained soils.
  \item ${\sigma'}_\mathrm{p}$ is dominated by $s_\mathrm{u}$, with
        water content $w$ and $LL$ entering as comparably strong
        contributors; Figure~\ref{fig:shap_scatter} shows that higher
        $s_\mathrm{u}$ increases the SHAP contribution to
        ${\sigma'}_\mathrm{p}$.
  \item $C_\mathrm{c}$ and $C_\mathrm{v}$ are reciprocally
        attributed to each other, consistent with the empirical
        coupling between compressibility and consolidation behaviour;
        $C_\mathrm{v}$, however, also depends on permeability, so this
        reciprocal pattern should not be read as a one-to-one physical
        mapping~\citep{Saito2025MIDM}.
\end{itemize}
Index properties nevertheless remain important secondary---and in
some cases near-co-dominant---contributors. The prominence of $LL$ for
$C_\mathrm{c}$ is physically consistent with the Skempton correlation
$C_\mathrm{c}\!\approx\!0.007(LL\!-\!10)$~\citep{Skempton1944}, while
the negative SHAP trend of ${\sigma'}_\mathrm{p}$ with water content
$w$ is consistent with the empirical tendency for wetter, softer clays
to exhibit lower preconsolidation pressure.
This cross-parameter attribution pattern is consistent with the iterative
imputation scheme, in which progressively refined estimates of correlated
parameters are used as non-target mechanical covariates.

\subsection{Embedding-Based Borehole-Structure Diagnostic}
\label{sec:embedding_diagnostic}

The preceding analyses examined what TabPFN predicts
(Section~\ref{sec:posterior}) and which input features carry the
strongest attribution for each prediction (Section~\ref{sec:shap}).
A natural follow-up question is whether the internal representation
that supports these predictions---namely the contextual embedding
TabPFN forms when fit to the BID training context (i.e., the complete
benchmark training rows)---also encodes borehole-level structure
beyond the per-sample mechanical target.
We address this through a post-hoc diagnostic on
TabPFN's target-specific embeddings, presented strictly as a
diagnostic indication of borehole-level structure rather than as
evidence of a validated borehole-context imputation mechanism.
To avoid making the borehole-grouping diagnostic tautological,
the borehole identifier is withheld from the embedding inputs
(i.e., the six index properties and four non-target mechanical
parameters) and used only as an evaluation label for within-
versus across-borehole structure.

We extract the contextual embedding $\mathbf{z}_i \in \mathbb{R}^{192}$
produced by TabPFN for each sample $i$, separately for each of the five
mechanical targets $s_\mathrm{u}$, $E_{\mathrm{u}}$,
${\sigma'}_\mathrm{p}$, $C_\mathrm{c}$, and $C_\mathrm{v}$,
using the BM/AirportSoilProperties benchmark context.
Because raw cosine similarities in the 192-dimensional embeddings are
concentrated near unity, we use two complementary diagnostics rather
than interpreting raw cosine values directly:
\begin{enumerate}
  \item \textbf{$k$-nearest-neighbour borehole purity}: for each sample,
        we compute the fraction of $k\!\in\!\{3,5,10,20\}$ Euclidean
        nearest neighbours that share its borehole identifier, and
        compare with a random-neighbour baseline (the per-sample analytic
        expectation $(n_b\!-\!1)/(n\!-\!1)$ for a borehole of size $n_b$).
  \item \textbf{PCA-reduced cosine similarity}: we project $\mathbf{z}_i$
        onto the leading $d \in \{2, 5, 10, 20, 50\}$ principal
        components and report
        $\Delta_\mathrm{cos}(d) = \overline{\cos}_\mathrm{within} -
         \overline{\cos}_\mathrm{across}$, the difference of mean cosine
        similarity between within-borehole and across-borehole sample
        pairs.
\end{enumerate}

\begin{figure}[htbp]
  \centering
  \includegraphics[width=\linewidth]{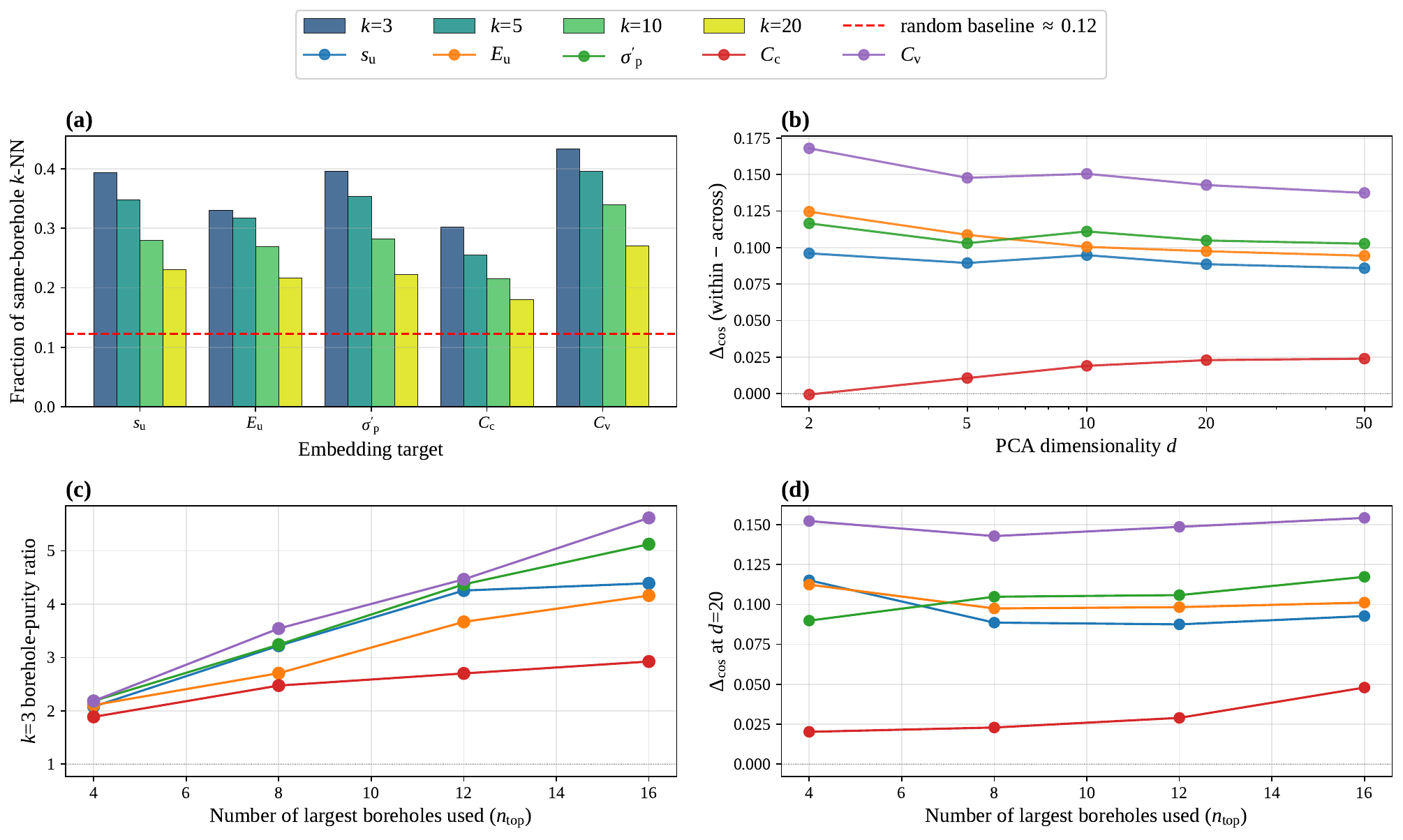}
  \caption{Embedding-based borehole-structure diagnostic for the five
           BM/AirportSoilProperties mechanical targets.
           (a) $k\!=\!3,\,5,\,10,\,20$ nearest-neighbour borehole purity per
           target at the default $n_\mathrm{top}\!=\!8$; the dashed line is
           the random-neighbour baseline.
           (b) PCA-reduced cosine $\Delta_\mathrm{cos}$ at $n_\mathrm{top}\!=\!8$
           as a function of dimensionality $d$.
           (c) $k\!=\!3$ borehole-purity ratio (mean purity divided by the
           random-neighbour baseline) against
           $n_\mathrm{top}\!\in\!\{4,8,12,16\}$; the dotted line marks
           ratio~$=\!1$ (chance level).
           (d) PCA-reduced cosine $\Delta_\mathrm{cos}$ at $d\!=\!20$ against
           $n_\mathrm{top}$.}
  \label{fig:embedding_diagnostic}
\end{figure}

The diagnostic is computed on the $n_\mathrm{top}\!=\!8$ largest
boreholes of the BID context ($270$ of the $2766$ BID rows, with
$32$--$38$ samples per borehole), because boreholes with only a few
samples yield too few within-borehole pairs to stably estimate the
within-versus-across similarity contrast.
Across all five mechanical targets, the $k\!=\!3$ nearest-neighbour
borehole purity exceeds the random baseline by a factor of $2.47$ to
$3.54$ (Figure~\ref{fig:embedding_diagnostic}a), and the PCA-reduced
cosine similarity is positive at $d\!=\!20$ in the range
$\Delta_\mathrm{cos}(20)\!\in\![+0.023,\,+0.143]$
(Figure~\ref{fig:embedding_diagnostic}b). Repeating the analysis for
$n_\mathrm{top}\!\in\!\{4, 8, 12, 16\}$ largest-borehole subsets indicates
that the result is not specific to the $n_\mathrm{top}\!=\!8$ choice:
all $5$ targets~$\times$~$4$ subsets~$=\!20$ combinations remain above
the random baseline (Figure~\ref{fig:embedding_diagnostic}c,d).
Among the five targets, the $C_\mathrm{c}$ embedding shows the weakest
borehole-grouping signal in both diagnostics, although it remains above
the random-neighbour baseline.

Beyond conventional site-level grouping, recent work argues that
parameter variability may not be fully captured at the geographic-site
level alone \citep{Saito2025Site}, motivating quasi-site-specific and
cluster-based inference approaches
\citep{Phoon2022,Phoon2025FR,Ching2021,Wu2022,Cai2024,Cai2025}. The
diagnostic above suggests that TabPFN-derived embeddings can, on this
benchmark, be inspected post-hoc to reveal that borehole-level grouping
remains visible in the embedding space---suggesting a possible
diagnostic entry point for such finer-grained schemes, although the
practical utility of this signal for downstream inference is not
validated here and is discussed further in Section~\ref{sec:discussion}.
We describe the result strictly as a borehole-level grouping signal:
the BM/AirportSoilProperties benchmark provides only borehole
identifiers, without depth or coordinate metadata, and we therefore do
not assert any geometric or distance-based interpretation of this
signal. The diagnostic also does not separate structure newly induced
by TabPFN from borehole-level grouping already present in the observed
index and non-target mechanical features; it only verifies that such
grouping remains visible in the learned embedding.

\subsection{Comparison with Baselines}
\label{sec:baseline_comparison}

To benchmark TabPFN's predictive accuracy, we compared it against
five reference baselines on the BM/AirportSoilProperties benchmark:
(i)~mean imputation, (ii)~linear
regression, (iii)~random forests, and (iv)~XGBoost---each run under
the same iterative-imputation procedure used in
Section~\ref{sec:iteration}---and (v)~the hierarchical Bayesian model
(HBM)~\citep{Ching2021, Wu2022} ($21{,}000$ Gibbs samples). The HBM
posterior summary obtained here is re-used in
Section~\ref{sec:rbd:settlement}.

RMSE on the official 20-sample missing pattern is reported in
Table~\ref{tab:baseline_comparison}; for the iterative imputers this
is the final-iteration RMSE, while the HBM value is computed from the
posterior mean.
TabPFN attains the lowest RMSE on four of the five mechanical targets
($s_\mathrm{u}$, ${\sigma'}_\mathrm{p}$, $C_\mathrm{c}$, $C_\mathrm{v}$);
on $E_{\mathrm{u}}$, XGBoost yields the lowest RMSE
($2376$~kN/m$^2$), HBM is the second-best ($2588$~kN/m$^2$), and
TabPFN is third ($2692$~kN/m$^2$); this is discussed further in
Section~\ref{sec:discussion}. We do not claim uniform superiority over
these baselines on this benchmark; we present TabPFN as competitive,
with the strongest performance on four of five targets while
additionally offering the embedding-similarity diagnostics of
Section~\ref{sec:embedding} and the predictive-distribution outputs
discussed in Section~\ref{sec:iteration}.

\begin{table}[htbp]
  \caption{Final-iteration RMSE on the BM/AirportSoilProperties
           20-sample missing pattern. Units follow each target.
           Lowest value per target is in bold.}
  \label{tab:baseline_comparison}
  \centering
  \begin{tabular}{lcccccc}
    \toprule
    & Mean imp. & Linear & HBM & RF & XGBoost & TabPFN \\
    \midrule
    $C_\mathrm{c}$                           & 0.332 & 0.127 & 0.128 & 0.108 & 0.123 & \textbf{0.106} \\
    $E_{\mathrm{u}}$ (kN/m$^2$)              &  3834 &  2952 &  2588 &  3212 & \textbf{2376} &  2692 \\
    $s_\mathrm{u}$ (kN/m$^2$)                & 24.50 & 16.19 &  9.74 & 14.64 & 16.69 & \textbf{9.30} \\
    ${\sigma'}_\mathrm{p}$ (kN/m$^2$)        & 49.93 & 35.44 & 28.24 & 32.92 & 38.79 & \textbf{25.89} \\
    $C_\mathrm{v}$ (m$^2$/s)                 & 168.71 & 119.16 & 59.32 & 46.91 & 60.90 & \textbf{34.58} \\
    \bottomrule
  \end{tabular}
\end{table}

\section{Reliability-Based Design Implication}
\label{sec:rbd}

This section continues the BM/AirportSoilProperties
regression analysis of Section~\ref{sec:regression}. Specifically, the
iteration-10 predictive distributions of the five mechanical parameters
(Section~\ref{sec:posterior}) serve as the input to the reliability
analyses below, rather than introducing a new dataset.

Linking these predictive distributions to reliability-based
design (RBD) considerations, and distinguishing the model's epistemic
uncertainty from the inherent aleatory variability of the soil
parameters, are two questions of direct relevance to engineering
practice. We explore both below in an
illustrative manner: Section~\ref{sec:rbd:proxy}
introduces a proxy decomposition of TabPFN's predictive uncertainty,
and Section~\ref{sec:rbd:settlement} couples TabPFN predictive distributions with a
one-dimensional consolidation settlement model to obtain a per-sample
reliability index $\beta$ and serviceability exceedance probability
$P_\mathrm{f}$.

\subsection{Proxy Decomposition of Predictive Uncertainty}
\label{sec:rbd:proxy}

For each missing sample $i$ and target parameter $y$, TabPFN returns a
predictive posterior $p(y_i \mid \mathcal{D}_\mathrm{obs})$ given the
observed BID context $\mathcal{D}_\mathrm{obs}$. We do not attempt a
strict epistemic--aleatory partition---which is generally not identifiable
without additional model assumptions
\citep{Hullermeier2021}---but instead report the
following additive proxy for predictive-variance contributions:
\begin{equation}
  \mathrm{Var}^{\mathrm{proxy}}\!\left[y_i \mid \mathcal{D}_\mathrm{obs}\right]
  \;\coloneqq\;
  \underbrace{\sum_{c \in \mathcal{C}}\mathrm{Var}_{s \in c}\!\bigl(\mathbb{E}[y_i \mid s]\bigr)}_{\text{scenario-sensitive (reducible) proxy}}
  \;+\;
  \underbrace{\frac{1}{|\mathcal{C}|}\sum_{c \in \mathcal{C}}\mathbb{E}_{s \in c}\!\bigl(\mathrm{Var}[y_i \mid s]\bigr)}_{\text{within-posterior predictive component}},
  \label{eq:proxy_decomp}
\end{equation}
where $\mathcal{C}$ is the scenario-class set defined below and $s$
indexes scenarios within a class. Equation~\eqref{eq:proxy_decomp}
defines
$\mathrm{Var}^{\mathrm{proxy}}[y_i\!\mid\!\mathcal{D}_\mathrm{obs}]$
as a constructed quantity (denoted by ``$\coloneqq$''); this proxy approximates the true predictive variance,
\begin{equation*}
  \mathrm{Var}^{\mathrm{proxy}}[y_i \mid \mathcal{D}_\mathrm{obs}]
  \;\approx\;
  \mathrm{Var}[y_i \mid \mathcal{D}_\mathrm{obs}],
\end{equation*}
with equality holding only when the scenario classes act independently on the predictive mean.
Because the
additivity of class-wise variances is an assumption rather than an
identity, we treat Eq.~\eqref{eq:proxy_decomp} as an additive proxy for
predictive-variance contributions, not as an identifiable variance
decomposition. The scenario-class set $\mathcal{C}$ contains four perturbation sources---three context-perturbation classes and one multi-seed run-to-run class---that we deliberately keep separate:
\begin{itemize}
  \item \textbf{Context bootstrap with replacement (WR)} — classical
        bootstrap resampling of the BID context. Rows may be duplicated;
        the expected distinct-row coverage is reduced both by row-level
        omission and by duplicate weighting.
  \item \textbf{Context bootstrap without replacement (WOR)} —
        row-level subsampling of the BID context without duplication. This
        class isolates the effect of distinct-row coverage loss from the
        duplicate-weighting effect of WR.
  \item \textbf{Context borehole-level holdout} — removal of a fraction
        of unique borehole identifiers from the BID context, so that
        entire boreholes are excluded as groups rather than as
        individual rows.
  \item \textbf{Multi-seed run-to-run variability} — twenty different
        random seeds passed to TabPFN while holding the BID context fixed;
        this component measures variation of the posterior mean across
        runs, not the within-posterior variance returned by TabPFN.
\end{itemize}
To compare the three context-perturbation classes on a single
information-content scale, we define the expected distinct-row coverage
$\kappa$ as the expected fraction of unique BID rows retained in the
perturbed context:
\begin{equation}
  \kappa_\mathrm{WR}(r) = 1 - (1 - 1/N)^{rN}, \quad
  \kappa_\mathrm{WOR}(r) = r, \quad
  \kappa_\mathrm{holdout}(f_h) \approx 1 - f_h,
  \label{eq:kappa}
\end{equation}
where $N = |\mathcal{D}_\mathrm{obs}^\mathrm{BID}| = 2766$, $r$ is the
bootstrap resample ratio, and $f_h$ is the held-out borehole fraction.
Each context class is evaluated on the grid
$\kappa \in \{0.3, 0.4, 0.5, 0.6, 0.7, 0.8, 0.9\}$ with $20$ replicates per
cell; the multi-seed class is run at $\kappa = 1.0$ as a baseline. For
the $\kappa$ sweep, we do not add the WR, WOR, holdout, and multi-seed
terms into one total. Instead, each plotted cell isolates one scenario class, so the
relevant per-cell total is
$\mathrm{Var}_{s \in c}(\mathbb{E}[y_i\!\mid\!s]) +
\mathbb{E}_{s \in c}(\mathrm{Var}[y_i\!\mid\!s])$ for that class alone.
We therefore report the across-scenario term separately for each class
and plot the within-posterior term as the average across the three
context classes at each $\kappa$. Point estimates carry $95\%$
percentile-bootstrap confidence intervals over the replicate index.

\begin{figure}[htbp]
  \centering
  \includegraphics[width=\linewidth]{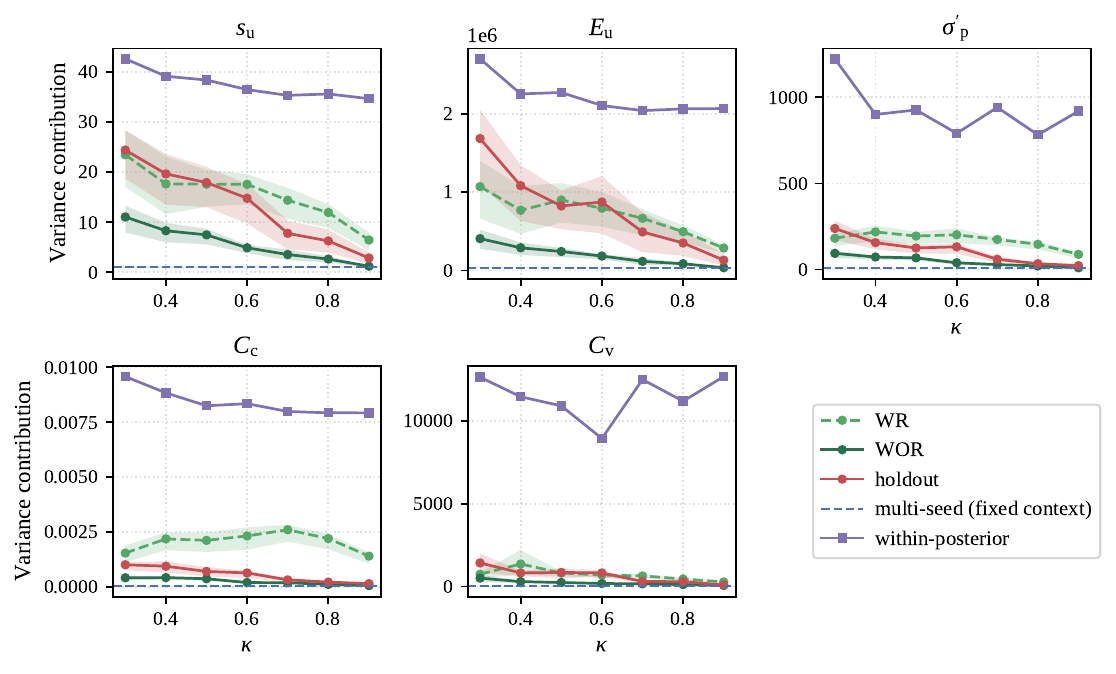}
  \caption{Matched-$\kappa$ proxy decomposition of TabPFN's
           predictive variance on the BM/AirportSoilProperties 20-sample
           missing pattern. Coloured curves are the across-scenario
           components for three context perturbations against $\kappa$:
           bootstrap with replacement (WR; dashed), without replacement
           (WOR; solid), and borehole-level holdout (holdout; solid), with
           $95\%$ percentile-bootstrap CI bands from $20$ scenarios per
           cell. The dashed blue baseline is the multi-seed
           run-to-run component estimated at $\kappa = 1.0$ with the
           context fixed. The violet line is the within-posterior
           component averaged across the three context classes; no CI
           band is shown for this cross-class average.}
  \label{fig:uncertainty_proxy_decomp}
\end{figure}

Three results are robust across the $\kappa$ grid. First, the
within-posterior predictive component dominates in every
$(\kappa,\mathrm{mode},\mathrm{target})$ cell: its fraction of the
single-class proxy total ranges from $0.61$ to $0.998$ with median
$0.91$ (Figure~\ref{fig:uncertainty_proxy_decomp}). Second, context
perturbations are appreciably larger than multi-seed run-to-run
variability; at $\kappa = 0.8$, the context-class variance exceeds the
multi-seed baseline by factors of $\times 2.4$ to $\times 44$ across
the five mechanical targets, with non-overlapping marginal $95\%$ CIs
in all $15$ $(\mathrm{target},\mathrm{mode})$ cells. Third, WOR is
consistently the smallest context-perturbation class at matched
$\kappa$: both $\mathrm{holdout} > \mathrm{WOR}$ and $\mathrm{WR} >
\mathrm{WOR}$ hold in $35/35$ $(\kappa,\mathrm{target})$ cells
($33/35$ and $34/35$ with non-overlapping marginal $95\%$ CIs,
respectively). By contrast, the ordering between WR and holdout is
mechanism- and $\kappa$-dependent ($25/35$ cells favour
$\mathrm{WR}$, with non-overlapping marginal $95\%$ CIs in only
$17/35$), so we do not present a universal rank order across all three
context classes. This residual ordering is
consistent with a trade-off between borehole-level coverage loss
(only present in holdout) and duplicate-weighting (only present in WR),
neither of which is present in WOR; we do not attempt a mechanistic
identification beyond this descriptive observation.

The $\kappa$-axis is our primary comparison because it matches
the expected number of distinct training rows, i.e., the amount of
unique information provided to TabPFN. A complementary sweep matching
the total number of context rows instead---under which WR has lower
distinct coverage because duplicates occupy part of the context
budget---reproduces the same within-posterior dominance,
``context exceeds multi-seed,'' and WOR-smallest findings. Because WR
decouples distinct rows from total rows, no single matching axis is
universally fair, and we therefore treat the WR-vs-holdout ordering
descriptively rather than as a universal rank.

A further qualification concerns the data requirements of the
procedure itself. The proxy decomposition reported here is a
data-rich demonstration: the context perturbations are repeatedly
applied to a comparatively abundant BID context ($N = 2766$), so that
even at the lowest coverage level examined ($\kappa = 0.3$) each
perturbed context still retains roughly $830$ distinct rows in
expectation. In genuinely data-scarce settings---for example, a site
investigation providing only tens of records from a few
boreholes---the perturbed contexts would become small, the resulting
across-scenario variance estimates could become unstable and
sensitive to individual records, and borehole-level holdout would
remove a large fraction of the available information at once. The
stability and usefulness of the proxy decomposition under such
limited-data conditions have not been established in this study, and
we treat its extension to genuinely data-scarce sites as an open
question.

\subsection{Illustrative Reliability Calculation for Settlement}
\label{sec:rbd:settlement}

To couple TabPFN's per-sample predictive distributions with a
reliability-style serviceability assessment, we propagate the marginal
predictive distributions of $C_\mathrm{c}$ and ${\sigma'}_\mathrm{p}$ through a
one-dimensional consolidation settlement model~\citep{Terzaghi1943}.
For a normally consolidated clay layer of thickness $H$ subject to a
uniform applied stress increment $\Delta\sigma$, the consolidation
settlement is
\begin{equation}
  S \;=\; \frac{C_\mathrm{c}}{1+e_0}\,H\,\log_{10}\!\frac{{\sigma'}_\mathrm{p}+\Delta\sigma}{{\sigma'}_\mathrm{p}},
  \label{eq:settlement_nc}
\end{equation}
where $e_0$ is the per-sample observed void ratio, treated as a
deterministic input, and ${\sigma'}_{v0}$ is taken as
${\sigma'}_\mathrm{p}$ (OCR~$=1$) because the benchmark lacks a
depth-resolved effective-stress profile. Secondary compression,
depth-varying stress, three-dimensional effects, and over-consolidated
behaviour requiring $C_\mathrm{r}$ lie outside the scope of this
illustrative calculation.

Drawing $N_\mathrm{MC}\!=\!20{,}000$ Monte Carlo samples independently
from the two per-sample marginal TabPFN predictive distributions, we compute the
serviceability exceedance probability
$P_\mathrm{f}\!=\!\mathrm{Pr}(S\!>\!S_\mathrm{allow})$ and reliability
index $\beta\!=\!-\Phi^{-1}(P_\mathrm{f})$, where $\Phi$ is the
standard normal cumulative distribution function. The illustrative
case uses $S_\mathrm{allow}\!=\!100\,\mathrm{mm}$,
$H\!=\!2\,\mathrm{m}$, and $\Delta\sigma\!=\!40\,\mathrm{kPa}$.
These design-case values are not code-calibrated parameters; they were
chosen as a hypothetical serviceability case for which the
deterministic settlement computed from the released benchmark truth
remains below $S_\mathrm{allow}$ for all $20$ official missing records
(maximum $75.3$~mm), while posterior uncertainty can still be propagated
into $P_\mathrm{f}$. Accordingly, we do not compare the resulting
$\beta$ values directly against design-code target values, as these
are calibrated for specific limit-state definitions outside the scope
of this illustrative model.

To gauge how strongly the reliability output depends on the posterior
source, we repeat the propagation using the HBM posterior summaries
from Section~\ref{sec:baseline_comparison}. For the two positive
parameters entering Eq.~\eqref{eq:settlement_nc},
$C_\mathrm{c}$ and ${\sigma'}_\mathrm{p}$, we represent each
per-sample marginal posterior by a log-normal distribution matched to
the available 5\%/95\% quantiles. This comparison is therefore
intended as an illustrative sensitivity check to the posterior source,
rather than as a dedicated joint HBM reliability analysis.

\begin{figure}[htbp]
  \centering
  \includegraphics[width=\linewidth]{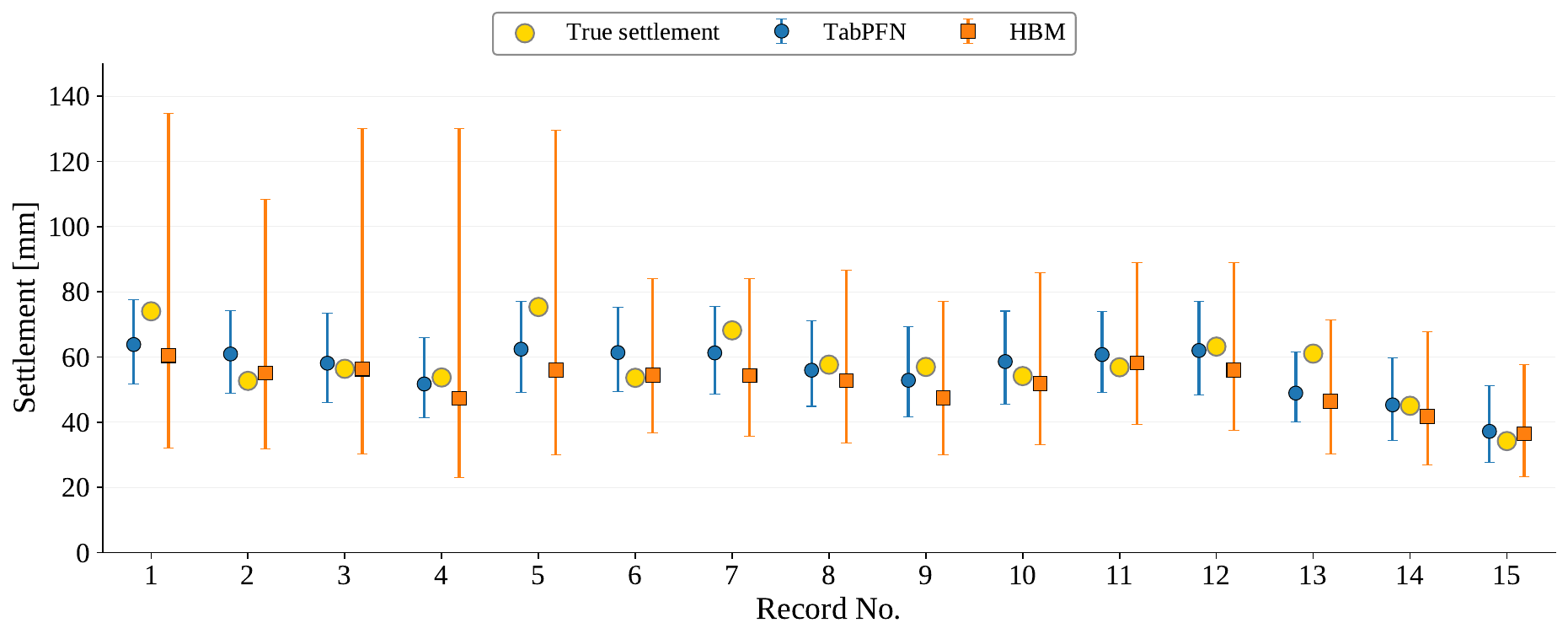}
  \caption{Per-sample settlement comparison between TabPFN (blue
           circles), HBM (orange squares), and the deterministic ground
           truth (gold circles) for the 15 jointly imputed samples.
           Markers show $S$ evaluated at the saved posterior mean of
           $(C_\mathrm{c}, {\sigma'}_\mathrm{p})$; error bars show
           5\%/95\% Monte Carlo intervals propagated through
           Eq.~\eqref{eq:settlement_nc}. HBM marginals are reconstructed
           by log-normal quantile matching.}
  \label{fig:rbd_truth_comparison_all}
\end{figure}

For the posterior-source comparison, $15$ of the $20$ samples have both
parameters missing and imputed by both methods; the remaining five
already contain both parameters in the missing-pattern row and are
excluded. Across these $15$ samples, the median $\beta$ is $2.22$ for
HBM, compared with $3.39$ for TabPFN. Using $S$
evaluated at the saved posterior mean of
$(C_\mathrm{c}, {\sigma'}_\mathrm{p})$ as the point prediction
(consistent with Table~\ref{tab:baseline_comparison}) and the
deterministic $S_\mathrm{true}$ computed from the benchmark's
released true parameters as the reference, the settlement RMSE is
$6.66$~mm for TabPFN and
$8.95$~mm for HBM (Figure~\ref{fig:rbd_truth_comparison_all}).
TabPFN's intervals are essentially constant ($25$--$27$~mm) across the three missing patterns, whereas HBM's central $90\%$ interval on $S$ narrows from $\sim\!97$~mm where neither $s_\mathrm{u}$ nor $E_\mathrm{u}$ is co-observed (records 1--5) to $\sim\!44$~mm where both are co-observed (records 11--15). The HBM behaviour is consistent with propagating additional uncertainty when correlated mechanical parameters are missing from the observation row, whereas TabPFN returns a context-conditioned predictive distribution whose width depends less strongly on which neighbouring targets are observed. Both methods cover $S_\mathrm{true}$ in their central $90\%$ intervals for all $15$ samples, and the resulting difference in median $\beta$ should therefore be interpreted as reflecting differing uncertainty-quantification approaches rather than a ranking of methods. Whether TabPFN's narrower intervals are well-calibrated under the missing patterns examined here remains a separate empirical question, and we treat the reliability indices obtained from either posterior source as illustrative.

We stress that the interval coverage reported above does not
constitute a calibration demonstration: with only $15$ samples,
central-interval coverage carries limited evidential weight, and it
does not test the tail behaviour on which $P_\mathrm{f}$ directly
depends. More fundamentally, the predictive distributions returned by
TabPFN quantify the model's in-context predictive uncertainty and
cannot be identified with a complete representation of the
underlying aleatory variability and epistemic uncertainty; a failure
probability obtained by propagating an
uncalibrated predictive distribution through a limit-state function
may therefore act as a black-box output that does not necessarily
reflect the inherent uncertainties of the project at hand.
Accordingly, the $P_\mathrm{f}$ and $\beta$ values reported here are
methodological illustrations of the propagation workflow, not
design-ready quantities, and they should not be used to support
design decisions in their present form. Site- and problem-specific
calibration assessment of the predictive distributions against an
independent and sufficiently representative local dataset, followed,
where necessary, by recalibration, should be carried out before
these outputs are used in operational reliability-based design. Such an
assessment and any necessary recalibration require data beyond those
available in the present study.

\section{Discussion}
\label{sec:discussion}

The examples studied here illustrate that TabPFN and its extensions
provide four post-prediction analysis routes---embedding similarity
(via \texttt{tabpfn-extensions}), posterior uncertainty
(via TabPFN's native quantised output), SHAP feature attribution
(via the permutation wrapper in \texttt{tabpfn-extensions}), and a
proxy decomposition of predictive uncertainty coupled with an
illustrative reliability-based design calculation
(Section~\ref{sec:rbd}). Across the
controlled classification example and the regression benchmark, the
model's outputs are consistent with established geotechnical expectations
without domain-specific fine-tuning, relying solely on in-context
learning from a small training set.

\paragraph{Embedding-based interpretability.}
The cosine-similarity heatmap (Figure~\ref{fig:embedding_heatmap})
shows label-consistent grouping in the controlled Clay/Sand example,
despite no explicit soil-type supervision at the embedding level.
This supports embedding inspection as a useful visual diagnostic in
this example; its use as a data-quality check for real borehole
records remains a validation target for future work.

\paragraph{Posterior uncertainty.}
The full posterior distributions obtained via TabPFN's quantised output
(Figure~\ref{fig:violin}) provide sample-specific uncertainty information
that can be propagated into the illustrative reliability calculation of
Section~\ref{sec:rbd}. Unlike point-estimate regressors, this output can
flag samples for which the prediction distribution is broad. In this
benchmark, $C_\mathrm{v}$---which is highly variable and difficult to
measure reliably---exhibits the broadest posteriors, while
$C_\mathrm{c}$ and ${\sigma'}_\mathrm{p}$ show comparatively narrow
distributions; this ordering is consistent with parameter-specific
predictability.

\paragraph{Proxy decomposition of predictive uncertainty.}
The $\kappa$-sweep decomposition (Section~\ref{sec:rbd:proxy};
Figure~\ref{fig:uncertainty_proxy_decomp}) supports two conclusions
that are robust across $\kappa$ and the five mechanical targets.
First, under the present in-context inference scheme, the single-class
proxy total is dominated by the within-posterior predictive
component---which we do not identify with the soil's aleatory
variability \citep{Hullermeier2021}---rather than by multi-seed
run-to-run variability; the context-perturbation terms exceed the
multi-seed baseline but remain substantially smaller than the
within-posterior component. Second, row-level subsampling without
replacement (WOR) is consistently the smallest of the three context
probes at matched $\kappa$, while the WR-vs-holdout ordering is
mechanism- and $\kappa$-dependent, consistent with a trade-off between
duplicate-weighting in WR and borehole-level coverage loss in holdout.

\paragraph{$E_{\mathrm{u}}$ and the limits of point-estimate iterative imputation.}
Among the five mechanical targets, $E_{\mathrm{u}}$ is the only one for
which TabPFN does not attain the lowest RMSE in
Section~\ref{sec:baseline_comparison}: XGBoost is best
(2376~kN/m$^2$), HBM is second (2588~kN/m$^2$), and TabPFN is third
(2692~kN/m$^2$). Per-pattern disaggregation shows that this gap is
concentrated in the all-mechanical-missing pattern (mt1; $n=5$), where
the TabPFN bias is approximately $-2{,}750$~kN/m$^2$ and the true mean
($\sim 9{,}000$~kN/m$^2$) lies in the BID upper tail, well above the
BID median of $\sim 4{,}300$~kN/m$^2$. This pattern is consistent with
a scheme-level limitation of the present deterministic iterative
imputation: when no mechanical covariates are observed, the missing
co-features enter the first pass through deterministic plug-in values
(initially BID means, with within-iteration sequential updates) and are
refined only deterministically thereafter, so co-feature uncertainty is
not propagated into the target prediction.
For heavy-tailed targets such as $E_{\mathrm{u}}$, this point-estimate
plug-in can pull the target prediction toward the population-mean
region, consistent with the observed mt1 underprediction. HBM
partially avoids this issue because Gibbs sampling stochastically
samples missing entries at each iteration, propagating co-feature
uncertainty to the target. We note, however, that we have not directly
demonstrated that posterior-sampling multiple imputation closes this
gap, and model-class effects (e.g., tree-based inductive bias suited
to heavy-tailed targets) cannot be ruled out as a partial contributor.
A natural future direction is to extend the present
iterative procedure into multiple imputation that samples
from TabPFN's predictive posterior at each step---a generalisation
that retains the underlying TabPFN model unchanged.

\paragraph{Feature importance and inter-parameter dependency.}
SHAP-based importance (Figures~\ref{fig:shap_bar} and \ref{fig:shap_scatter})
shows a two-regime attribution pattern in the inter-parameter
dependency 
captured by the model.
For consolidation parameters, mutual cross-parameter influence and index
properties both play roles: $C_\mathrm{v}$ leads for $C_\mathrm{c}$ with $LL$
as the top index property (consistent with the Skempton correlation), while
$C_\mathrm{c}$ dominates $C_\mathrm{v}$ with $LL$ suppressing consolidation
rate in high-plasticity soils.
For strength and stiffness parameters ($s_\mathrm{u}$, $E_{\mathrm{u}}$,
${\sigma'}_\mathrm{p}$), other mechanical parameters outrank index properties, with
$w$ emerging as a 
co-dominant predictor of ${\sigma'}_\mathrm{p}$,
suggesting that the iterative imputation scheme 
propagates information across correlated targets.
This two-regime structure---index-property and cross-parameter influence for
compressibility, cross-parameter dominance for strength---is consistent with
established geotechnical knowledge; 
we read this as an associative attribution pattern that does not establish causal physical mechanisms.
Compared with hierarchical Bayesian approaches~\citep{Ching2021, Wu2022},
which characterise inter-parameter correlations through explicitly specified
probabilistic models, the SHAP-based analysis provides a complementary and
more granular view: it quantifies asymmetric, nonlinear attribution patterns between
individual parameters at the sample level, without requiring a parametric
correlation structure to be assumed in advance.

\paragraph{Comparison with existing approaches.}
Table~\ref{tab:comparison}
enumerates the post-prediction analysis tools natively provided by
TabPFN-extensions; hierarchical Bayesian models
(HBM)~\citep{Ching2021, Wu2022} and conventional machine-learning
methods such as random forests and recurrent neural
networks~\citep{Zhang2021RF, Manandhar2026} are included as reference
points. Accordingly, Table~\ref{tab:comparison} should not be read as
an overall ranking of modelling frameworks; it is a catalogue of
post-prediction tools available in the TabPFN-extension workflow.
The comparison focuses not on predictive accuracy, which
is
addressed separately in Section~\ref{sec:baseline_comparison} and in our
companion study~\citep{Saito2025}, but on the range of post-prediction
analysis tools available to the practitioner.
HBM provides model-based posterior distributions under explicit assumptions
on the prior and likelihood, but does not natively offer embedding-based
similarity analysis or model-agnostic feature attribution.
Conventional ML methods can be coupled with permutation-based SHAP, yet
they produce only point estimates unless additional techniques (e.g.\
quantile regression forests, Monte Carlo dropout) are employed, and lack
a built-in embedding representation.
TabPFN combines native posterior predictive outputs, contextual embeddings, and
SHAP attribution within a single framework, all without model retraining
or hyperparameter tuning.
In addition to the capabilities listed in Table~\ref{tab:comparison},
HBM offers further strengths such as transparent uncertainty decomposition;
conventional machine-learning methods offer tunable hyperparameters and
well-established theoretical foundations.

\begin{table}[htbp]
  \caption{
           TabPFN-centric capability summary;
           complementary strengths of HBM and conventional ML are
           discussed in the surrounding paragraph.}
  \label{tab:comparison}
  \centering
  \begingroup
  \small
  \setlength{\tabcolsep}{4pt}
  \begin{tabular}{@{}lccc@{}}
    \toprule
    Capability & HBM & Conv.\ ML$^{\dagger}$ & TabPFN + Ext. \\
    \midrule
    Posterior predictive distribution        & Yes (parametric) & No$^{\ast}$ & Yes (native) \\
    Embedding similarity          & Not native & No  & Yes \\
    SHAP feature importance       & Indirect$^{\ddagger}$ & Yes & Yes \\
    No retraining required        & No  & No  & Yes \\
    Correlation structure         & Parametric & --- & Implicit / prior-learned \\
    User-specified priors         & Yes & No & No \\
    Explicit Bayesian updating    & Yes & No & No \\
    \bottomrule
    \multicolumn{4}{@{}p{0.98\linewidth}@{}}{\footnotesize $^{\ast}$Requires additional techniques
    (e.g.\ quantile regression forests, Monte Carlo dropout).} \\
    \multicolumn{4}{@{}p{0.98\linewidth}@{}}{\footnotesize $^{\dagger}$Conventional ML methods
    (e.g.\ random forests, RNN)~\citep{Zhang2021RF, Manandhar2026}.} \\
    \multicolumn{4}{@{}p{0.98\linewidth}@{}}{\footnotesize $^{\ddagger}$Indirect: achievable via model-agnostic
    wrappers (e.g.\ KernelSHAP) or HBM-native posterior sensitivity analysis.}
  \end{tabular}
  \endgroup
\end{table}

\paragraph{Limitations.}
Several limitations should be acknowledged.
As an in-context learner, TabPFN's decision boundaries are governed entirely
by the training samples provided at inference time; in regions of the feature
space that lack training samples of one class, the model cannot infer a
physically meaningful boundary and may extrapolate in an unintuitive direction.
This is evident in the low-$N$, low-$V_\mathrm{s}$ region of
Figure~\ref{fig:proba_heatmap}, 
which illustrates that the
probability surface can behave counterintuitively in locally
class-sparse parts of the input space.
Expanding or diversifying the training set to cover the full feature space
would mitigate this behaviour.
Additionally, the iterative imputation procedure used for the regression
benchmark assumes that the model's predictive distribution is well-calibrated
at each iteration; systematic miscalibration could propagate errors across
targets.
Further validation on independent datasets from different geological settings
is needed to confirm the generalisability of the approach.

Several additional limitations apply to the new analyses of
Sections~\ref{sec:embedding_diagnostic} and~\ref{sec:baseline_comparison}. The embedding-based borehole-structure
diagnostic of Section~\ref{sec:embedding_diagnostic} reports a
consistent borehole-level grouping signal in TabPFN's target-specific
embeddings, but we have not validated whether this signal improves
imputation accuracy on the official $20$-sample missing pattern. We
therefore present this finding as an indication of site-related
structure, rather than as evidence that TabPFN exploits this signal
in a validated imputation mechanism. The baseline comparison in
Section~\ref{sec:baseline_comparison} is restricted to the official
$20$-sample missing pattern; we did not retrain the baseline models
under bootstrap or borehole-level holdout context perturbations, and
therefore do not extend the proxy decomposition to those models.

The reliability-related analyses of Section~\ref{sec:rbd}
introduce further caveats. The decomposition in
Section~\ref{sec:rbd:proxy} is a proxy: we do not identify
the within-posterior predictive component with the soil's aleatory
variability, since this equivalence is only approximate
\citep{Hullermeier2021}. The $\kappa$-axis sweep
(Figure~\ref{fig:uncertainty_proxy_decomp}) shows that the
within-posterior dominance and the ``context-perturbation variance
exceeds the multi-seed baseline'' findings are robust across
$\kappa \in [0.3, 0.9]$; the WR-vs-holdout ordering is mechanism- and
$\kappa$-dependent, and is therefore reported descriptively rather
than as a universal rank order across all three context classes. The illustrative reliability
calculation of Section~\ref{sec:rbd:settlement} uses a fixed design
case ($H\!=\!2\,\mathrm{m}$, $\Delta\sigma\!=\!40\,\mathrm{kPa}$) and a one-dimensional
consolidation model, with $e_0$ treated as a per-sample deterministic
input from the BID context. The TabPFN-vs-HBM comparison uses
marginal posterior summaries for $C_\mathrm{c}$ and
${\sigma'}_\mathrm{p}$ in both methods; spatial correlation across
boreholes, cross-parameter posterior dependence, and FORM/SORM-based
reliability analysis under correlated inputs are beyond the scope of
this study.

Consistent with these caveats, we regard site- and
problem-specific calibration assessment of the predictive
distributions, with recalibration where necessary, as an
important step that should precede operational reliability-based
design application of the workflow illustrated in
Section~\ref{sec:rbd}: failure probabilities
propagated from uncalibrated predictive distributions may act as
black-box outputs, and the $\beta$ and $P_\mathrm{f}$ values of
Section~\ref{sec:rbd:settlement} are methodological illustrations
rather than design-ready quantities. The proxy decomposition of
Section~\ref{sec:rbd:proxy} was likewise demonstrated only with a
comparatively data-abundant context ($N = 2766$); its stability and usefulness
under genuinely limited context data have not been established in
this study.

\section{Conclusion}
\label{sec:conclusion}

We have explored a set of interpretability and reliability-related tools
based on TabPFN and its extensions---embedding analysis
(via \texttt{tabpfn-extensions}), posterior distributions
(via TabPFN's native quantised output), SHAP-based feature importance
(via \texttt{tabpfn-extensions}), and a proxy decomposition of predictive
uncertainty coupled with an illustrative settlement-reliability calculation
(this study, Section~4)---that may be useful for selected geotechnical
analyses.
For the soil classification task, cosine-similarity heatmaps of learned
embeddings
showed label-consistent grouping of Clay and Sand samples without any explicit
soil-type supervision, 
in a way
that is consistent with the soil-type labels in this controlled
illustrative case.
For the regression benchmark, iterative imputation improved final RMSE
relative to iteration~1 for all five mechanical parameters, posterior
distributions showed parameter-specific widths (e.g., broad posteriors for
$C_\mathrm{v}$, narrow for $C_\mathrm{c}$), and SHAP analysis produced
attribution patterns consistent with established geotechnical relationships
such as the Skempton correlation and the inverse dependence of
${\sigma'}_\mathrm{p}$ on water content.
The proxy decomposition of predictive-variance contributions,
swept along a distinct-row coverage axis $\kappa$ and evaluated
separately for each context-perturbation class, shows that the
within-posterior predictive component is largest across all
configurations and that context perturbation exceeds multi-seed
run-to-run variability by an appreciable margin. Among context probes,
WOR is consistently the smallest at matched $\kappa$, whereas the
WR-vs-holdout ordering is mechanism- and $\kappa$-dependent and is
reported descriptively rather than as a universal rank.
Both reliability-related analyses carry scope qualifications:
the proxy decomposition was demonstrated only with a comparatively
data-abundant context, and its behaviour under genuinely limited
context data remains an open question; the settlement
calculation is a methodological illustration whose outputs would
require site- and problem-specific calibration assessment and, where
necessary, recalibration
before operational reliability-based design use
(Section~\ref{sec:rbd}).
These results suggest that TabPFN, when coupled with its extension tools, may serve as one of several
modern baselines for AI-assisted site characterisation under data-scarce
conditions, with comparable accuracy on the BM/AirportSoilProperties
benchmark and additional interpretability tools available.
Future work includes domain-specific fine-tuning of the foundation model on
large-scale geotechnical databases to develop a civil-engineering-specialised
variant, extending the deterministic iterative imputation of
Section~\ref{sec:regression} into a multiple-imputation scheme that
samples from TabPFN's predictive posterior at each step, as well as
exploration of large language model (LLM) integration for
natural-language explanation of predicted parameters and uncertainty estimates.

\begin{ack}
This research was supported by JSPS KAKENHI Grant Numbers JP23H00195 and JP25KJ0619.
\end{ack}

\bibliography{biblio.bib}

\end{document}